\documentclass[
 reprint,
superscriptaddress,
nofootinbib,
 amsmath,amssymb,
 aps,
 prd,
]{revtex4-1}

\usepackage{graphicx,color}
\graphicspath{{./figures/}}
\usepackage{dcolumn}
\usepackage{bm}
\usepackage{hyperref}
\usepackage{physics}
\usepackage{soul}
\usepackage{cancel}
\usepackage{mathtools}
\usepackage[normalem]{ulem}

\usepackage[caption=false]{subfig}

\begin{document}
\preprint{APS/123-QED}
\title{Jet-like structures in low-mass binary neutron star merger remnants}
\author{Jamie Bamber}
\affiliation{%
Department of Physics, University of Illinois at Urbana-Champaign, Urbana, IL 61801, USA
}
\author{Antonios Tsokaros}
\affiliation{%
Department of Physics, University of Illinois at Urbana-Champaign, Urbana, IL 61801, USA
}%
\affiliation{National Center for Supercomputing Applications, University of Illinois at Urbana-Champaign, Urbana, IL 61801, USA}
\affiliation{Research Center for Astronomy and Applied Mathematics, Academy of Athens, Athens 11527, Greece}
\author{Milton Ruiz}%
\affiliation{%
 Departament d’Astronomia i Astrof\'{i}sica, Universitat de Val\`{e}ncia, C/ Dr Moliner 50, 46100, Burjassot (Val\`{e}ncia), Spain
}
\author{Stuart L. Shapiro}
\affiliation{%
Department of Physics, University of Illinois at Urbana-Champaign, Urbana, IL 61801, USA
}%
\affiliation{
Department of Astronomy, University of Illinois at Urbana-Champaign, Urbana, IL 61801, USA
}
\affiliation{National Center for Supercomputing Applications, University of Illinois at Urbana-Champaign, Urbana, IL 61801, USA}

\date{\today}

\begin{abstract}
GW170817 and GRB 170817A provided direct evidence that binary neutron star (NSNS) mergers can produce short gamma-ray bursts (sGRBs). However, questions remain about the nature of the central engine. Depending on the mass, the remnant from a NSNS merger may promptly collapse to a black hole (BH), form a hypermassive neutron star (HMNS) which undergoes a delayed collapse to a BH, a supramassive neutron star (SMNS) with a much longer lifetime, or an indefinitely stable NS with a mass below the TOV limit. There is strong evidence that a BH with an accretion disk can launch a sGRB-compatible jet via the Blandford-Znajek mechanism, but whether a supramassive star can do the same is less clear. We have performed general relativistic magnetohydrodynamics simulations of the merger of both irrotational and spinning, equal-mass NSNSs constructed from a piecewise polytropic representation of the nuclear SLy equation of state, with a range of gravitational masses that yield remnants with mass above and below the supramassive limit. Each NS is endowed with a dipolar magnetic field extending from the interior into the exterior, as in a radio pulsar. We examine cases with different initial binary masses, including a case which produces a HMNS which collapses to a BH, and lower mass binaries that produce SMNS remnants. We find similar jetlike structures (helical magnetic field structures, a magnetically dominated evacuated funnel, and mildly relativistic outflow from the poles) for both the SMNS and HMNS remnants that meet our basic criteria for an \textit{incipient} jet. The outflow for the HMNS case is consistent with a Blandford-Znajek (BZ) jet. There is sufficient evidence that such BZ-powered outflows can break out and produce ulrarelativistic jets so that we can describe the HMNS system as a sGRB progenitor. However, the incipient jets from the SMNS remnants have much more baryon pollution and we see indications of inefficient outflow acceleration and mixing with the surrounding debris torus. Therefore, we cannot conclude that outflows from SMNSs are the progenitors of sGRBs.
\end{abstract}

\maketitle

\section{Introduction}
\label{sec:Intro}
The gravitational wave (GW) event GW170817 \cite{LIGOScientific:2017vwq} observed by the LIGO-Virgo consortium was remarkable both as the first GW signal identified as a NSNS merger, and because of the coincident observation of a short $\gamma$-ray burst (sGRB) GRB 170817A  \cite{LIGOScientific:2017zic,Coulter:2017wya,Savchenko:2017ffs,Lazzati:2016yxl,Murguia-Berthier:2017kkn,Duffell:2018iig,Lamb:2018ohw,Lamb:2018qfn,Mooley:2018qfh,Mooley:2018clx,Wu:2018bxg} followed by observations accross the electromagnetic spectrum. This ushered in the era of multimessenger [GW + electromagnetic (EM)] astronomy and provided direct evidence that sGRBs (or at least a subset of them) come from compact binary mergers where at least one of the companions is a neutron star (NS), as predicted by \cite{Paczynski:1986px,Goodman:1986az,Eichler:1989ve,Narayan:1992iy,Fox:2005kv,DAvanzo:2015kdp}. The total mass of the progenitor system of GW170817  is constrained to $(2.73,3.29)M_{\odot}$ with 90\% confidence. The individual masses $m_1,m_2$ of the binary components were inferred as $m_1 \in (1.36,2.26)M_{\odot}$ and $m_2 \in (0.86,1.36)M_{\odot}$, with the uncertainty being due to the degeneracy between the mass ratio and the aligned spin components \cite{LIGOScientific:2017vwq}. These values are within the observed range of NS masses \cite{Kiziltan:2013oja,Landry:2021hvl}, while below the lower limit of the BH mass distribution as inferred from X-ray binary observations and formation from stellar collapse \cite{Ozel:2010su,Kreidberg:2012ud}. While mechanisms have been proposed for the formation of BHs with smaller masses (see \cite{Yang:2017gfb,Carr:2020gox}), the observational evidence for BHs with masses below $3M_{\odot}$ is very weak, supporting the identification of GW170817 as a NSNS merger. 

sGRBs are characterised by a prompt emission of hard $\gamma$-rays with duration $T_{90} \lesssim 2$s, where $T_{90}$ is the time interval containing 90\% of the total $\gamma$-ray count (see \cite{Nakar:2007yr,Kumar:2014upa,DAvanzo:2015kdp,Piran:2004ba} for detailed reviews). GRB 170817A was identified \cite{Connaughton:2017,Goldstein:2017mmi,Savchenko:2017ffs} $1.734 \pm 0.054$s after GW170817 \cite{LIGOScientific:2017ync} with a duration $T_{90} = 2.0 \pm 0.5$s. An optical transient SSS17a / AT 2017gfo \cite{Coulter:2017wya} was observed $10.87$hrs after the GW signal in the host galaxy NGC 4993 at a distance $\sim 40$ Mpc, consistent with the $40\pm 8$ Mpc distance inferred from GW170818, with initial UV-blue dominated emission dimming and reddening to infrared over subsequent days. Subsequent observations also identified the source in x-ray and radio wavelengths (see e.g.~\cite{DAvanzo:2018zyz,Resmi:2018wuc}). 

The origin of the $\gamma$-ray emission in sGRBs is thought to be a narrowly collimated highly relativistic jet,~\footnote{where by ``jet" we mean a collimated outflow of EM fields and plasma along the rotational axis of the source \cite{DeYoung:1991}.} with Lorentz factors of at least $\Gamma \gtrsim 20$ \cite{Ghirlanda:2017opl} with typical values being $\Gamma \sim O(10^2)$ \cite{Granot:2005ye,Zou:2010,Paczynski:1986px,Ghirlanda:2017opl}. The typical isotropic-equivalent $\gamma$-ray luminosity of observed sGRBs is $\sim 10^{49}-10^{54}$ erg s$^{-1}$ \cite{Li:2016pes,Beniamini:2020adb,Shapiro:2017cny} corresponding to a real $\gamma$-ray luminosity of $10^{47}-10^{52}$ erg s$^{-1}$ \cite{Zhu:2023nkx}. The burst is characterized by hard prompt emission arising from either internal shocks between shells within the jet \cite{Rees:1994nw,Piran:2004ba} or external shocks from the leading shell \cite{Katz:1993fn}, at distances $\gtrsim 10^{5}$ km \cite{Veres:2018trt}.

It has been suggested that the additional soft thermal $\gamma$-ray component of the GRB 170817 emission originates from the hot dense cocoon surrounding the jet \cite{Goldstein:2017mmi}. The unusually low isotropic-equivalent luminosity of GRB 170817 ($L_{\gamma,\textup{iso}} \sim 10^{47}$erg s$^{-1}$), the spectral lag of the afterglow, and radio emission consistent with superluminal apparent motion \cite{Fraija:2017aev,Mooley:2018qfh}, have been attributed to its jet being viewed $\sim 20-30^{\circ}$ off-axis with a half-opening angle of the jet core of $\lesssim 5^{\circ}$ (see e.g \cite{Ghirlanda:2018uyx,Troja:2020pzf}).
Indeed it has been suggested that the $\gamma$-ray flux we observed from GRB 170817 is from a sheath of slower material surrounding the jet core rather than the core itself due to the oblique viewing angle \cite{Ghirlanda:2018uyx}, while the on-axis emission from the core likely has an isotropic-equivalent luminosity of $\geq 10^{51}$ erg s$^{-1}$, similar to other sGRBs, powered by a isotropic-equivalent kinetic energy outflow of $\gtrsim 10^{53}$ erg s$^{-1}$. The late time non-thermal $X$-ray and radio afterglow has been attributed to the interaction with the interstellar medium and the production of synchrotron radiation at the external forward-shock \cite{Fraija:2017aev}, while the UV/optical/infrared transient is consistent with the kilonova or macronova model \cite{Cowperthwaite:2017dyu,Smartt:2017fuw,Kasliwal:2017ngb}: a thermal mostly-isotropic transient powered by the radioactive decay of unstable nuclei formed from rapid neutron capture (the r-process) in the neutron-rich non-relativistic ejecta (see \cite{Metzger:2019zeh} for a review). The ejecta mass has been estimated as $0.04 \pm 0.01 M_{\odot}$ ($\sim 1.4\%$ of the total binary mass) with velocities of $\sim 0.1c$ and $\sim 0.3c$ for the red and blue components respectively.  

While the estimated masses of the binary companions strongly suggest that GW170817 represents a NSNS merger, its post-merger fate is uncertain. In principle, a NSNS merger can produce one of four possible outcomes depending on the equation of state (EOS), the mass, and the spin of the post-merger remnant \cite{Shibata:2006nm}: 
i) If the mass is below the Tolman–Oppenheimer–Volkoff (TOV) limit, the maximum mass $M_{\textup{TOV}}$ for a zero-temperature nonrotating NS, then the merger remnant will live for a very long-time as a spinning NS. In the presence of a dissipative process, 
e.g. from pulsar magnetic dipole emission, angular momentum is removed, and the NS ultimately spins down 
to a nonrotating stable NS. The spindown timescale is $\sim 10^{3}B^{-2}_{15}\;T^2_{\textup{ms}}\;\textup{s}$ for magnetic dipole radiation \cite{Shapiro:1983,Thompson:2004wi,Lander:2018und} where $B_{15}$ is the magnetic field strength in units of $10^{15}G$ and $T_{\textup{ms}}$ the rotation period in ms. ii) Remnants with mass larger than $M_{\textup{TOV}}$ but smaller than the maximum mass for a uniformly rotating zero-temperature NS, $M_{\textup{sup}}$, are termed \textit{supramassive} neutron stars (SMNSs) \cite{Cook92b} (see Fig. \ref{fig:Mrho}).  
Similar to (i), in the presence of a dissipative process, SMNSs also spin down, but this time the endpoint of their evolution is a BH instead of stable NS \cite{Friedman:1988er}. The SMNS lifetime depends similarly on $B$ and $T$, and is typically of the order $\sim 10^3$s \cite{Beniamini:2021tpy}, but its exact value depends on how close the star is to the turning point for uniformly rotating stars. 
iii) For masses larger than $M_{\textup{sup}}$ a metastable \textit{hypermassive} neutron star (HMNS) forms that can be supported only by differential rotation \cite{Baumgarte:1999cq}. The hypermassive star persists for many orbital periods, typically $O(10)$ms, before collapsing to a BH (see e.g.~\cite{Duez:2004nf,Ruiz:2017due}), after the differential rotation is lost through viscous effects, magnetic winding, and GWs. 
iv) Finally, for total initial binary masses above some dynamically determined threshold, $M_{\textup{thresh}}$, which depends on the EOS and the initial NS spin, the remnant undergoes prompt collapse to BH on a timescale of only a few ms \cite{Koppel:2019pys,Kolsch:2021lub}.

A key open question in ascertaining the fate of the merger in event GW170817, as well as the central engine behind sGRBs, is whether an ultarelativistic jet can only be powered by a BH with an accretion disk, or whether it can also be powered by a long-lived, highly magnetized NS remnant immersed into a gaseous environment of tidal debris \cite{Usov:1992zd,Duncan:1992hi,Zhang:2000wx,Metzger:2007cd,Metzger:2011,Bucciantini:2011kx,Metzger:2013cha,Ruiz:2017due,Ciolfi:2020hgg}. Accretion onto a spinning BH can power an ultrarelativistic jet through either the Blandford-Znajek (BZ) mechanism \cite{Blandford:1977ds}, neutrino-antineutrino annihilation along the BH spin axis \cite{Eichler:1989ve,Ruffert:1998vp}, or a combination of both, although the BZ mechanism is more likely to produce the luminosity consistent with observed sGRBs \cite{Kyutoku:2017voj}. BHs that form from the prompt collapse of merger remnants are unlikely to produce jets, as there is not enough time for the magnetic field to grow to force-free values above the BH poles \cite{Ruiz:2017inq}, a requirement of the BZ mechanism. On the other hand, the more highly magnetized accretion disk that forms around the BH after the collapse of the HMNS creates the optimal conditions for a BZ-driven jet \cite{Ruiz:2016rai,Usov:1992zd}.  

Jet formation from NS remnants however remains an open problem. These NS remnants generally do not have the required ergosphere for the BZ mechanism,\footnote{However, see \cite{Tsokaros:2019mlz,Tsokaros:2020qju} for extreme examples of NSs with an ergosphere.} however the large reservoir of rotational energy of the star ($\sim 10^{53} \textup{erg}$) can in principle be sufficient to power a sGRB, if it can be efficiently extracted via magnetic processes \cite{Usov1992,Ciolfi:2018tal,Ciolfi:2019fie,Bucciantini:2011kx}. Slowly decaying ``X-ray plateaus", lasting $10^2-10^5$s, in the soft X-ray afterglow of a subset of sGRBs have been cited as evidence for continuous energy injection from a magnetar central engine \cite{Ciolfi:2018tal,Rowlinson:2013ue,Zhang:2000wx} (although no such plateau was observed in the afterglow of GRB 170817 \cite{LIGOScientific:2017zic,Margalit:2017dij}). The timescales are significantly larger than the accretion timescale for a stellar mass BH, while the X-ray emission could be explained by spin down radiation from a NS~\cite{Jordana-Mitjans:2022gxy}. However, several alternative explanations (e.g. \cite{Strang:2019piq,Oganesyan:2020,Dereli-Begue:2022clf,Ciolfi:2014yla}) for these features have been proposed which are compatible with the BH + disk model. The additional energy injection from a magnetar remnant increases the energy of the quasi-isotropic ejecta and the associated kilonova, producing stronger radio emission at late-times \cite{Metzger:2013cha}. The non-detection of such late-time radio emission has been used to rule out a magnetar remnant central engine for some sGRBs \cite{Metzger:2013cha,Horesh:2016dah,Fong:2016orv}. Conversely, other authors have argued that, given the uncertainties in the physical parameters, radio observations of sGRB afterglows remain broadly compatible with magnetar remnants \cite{Liu:2019fgl}.

The main challenge for the magnetar central engine model is the requirement for a relatively baryon-free environment to launch the jet. Neutrino radiation from the NS remnant and / or magnetic processes produce an isotropic baryon wind in addition to the dynamical ejecta \cite{Dessart:2008zd,Siegel:2017nub}, and the resulting baryon pollution may limit the maximum terminal Lorentz factor of the jet to $\mathcal{O}(10)$ \cite{Sarin:2020gxb}, less than the typical $\gtrsim 100$ which is required. It has also been argued that if a jet does launch from a magnetar remnant it needs to do so within $\leq 100$ms post-merger to avoid the jet becoming choked by the wind \cite{Murguia-Berthier:2014pta}. 

Several different groups have conducted general relativistic magnetohydrodynamic (GRMHD) simulations of NSNS mergers over the last two decades (for instance \cite{Price:2006fi,Duez:2006qe,Anderson:2008zp,Giacomazzo:2010bx,Rezzolla:2011da,Neilsen:2014hha,Palenzuela:2015dqa,Kiuchi:2015sga,Sekiguchi:2015dma,Ruiz:2016rai,Kawamura:2016nmk,Kiuchi:2017zzg,Dietrich:2017xqb,Ruiz:2019ezy,Ciolfi:2019fie,Ruiz:2020via,Ciolfi:2020hgg,Mosta:2020hlh,Ruiz:2021qmm,Sun:2022vri,Kiuchi:2022nin,Zappa:2023,Foucart:2022kon,Most:2023sft,Radice:2023zlw,Combi:2023yav,Kiuchi:2023obe,Aguilera-Miret:2023qih,Zenati:2024pgn}) each with different strengths and weaknesses. Simulations performed by our group (e.g. \cite{Ruiz:2016rai}) demonstrated the formation of a collimated, mildly relativistic outflow, with a tightly wound helical magnetic field from the poles, powered by the BZ process from a BH with a magnetized accretion disk formed following the collapse of a HMNS remnant \cite{Ruiz:2016rai}. We identified this outflow as an {\it incipient jet}. The NSs were irrotational, equal mass and modeled with a polytropic $\Gamma=2$ EOS, where $\Gamma$ is the adiabatic index. Similar results were obtained from simulations of black hole-neutron star (BHNS) mergers \cite{Paschalidis:2014qra,Ruiz:2018wah} which also result in a BH surrounded by a magnetized accretion disk. Our studies were later followed up with simulations where the stars had initial spin \cite{Ruiz:2019ezy}, different orientations of the magnetic dipole moment \cite{Ruiz:2020via}, realistic piecewise polytropic EOSs (SLy and H4) \cite{Ruiz:2021qmm}, and simulations incorporating a M1 neutrino transport scheme \cite{Sun:2022vri}. These studies showed:
\begin{enumerate}
\item The larger the spin of the progenitor stars, the heavier the disk (the smaller the mass of the BH remnant), and the shorter the delay time before a jet is launched (following the collapse of the HMNS remnant). NSNS with aligned spins enhance the magnetic field amplification (following merger) more efficiently than the irrotational ones \cite{Ruiz:2019ezy}. 
\item An incipient jet emerges whenever there is a sufficiently large poloidal component of the initial magnetic field aligned with the orbital angular momentum axis. The lifetime $\Delta t \gtrsim 140 (M_{\rm NS}/1.625M_\odot)$ ms and EM luminosity $L_{\textup{EM}} \sim 10^{52\pm1}$ erg s$^{-1}$ were consistent with typical sGRBs, as well as with the BZ mechanism \cite{Ruiz:2020via,Shapiro:2004ud}.
\item The softer the EOS, the larger the amount of matter ejected following the NSNS merger. The ejecta can be up
to a factor of $\sim 8$ larger in magnetized NSNS mergers than that in unmagnetized ones \cite{Ruiz:2021qmm}.
\item The inclusion of neutrino radiation \cite{Sun:2022vri} was found to induce an additional effective viscosity allowing for further angular momentum transport and faster collapse of the HMNS remnant to a BH. Magnetic fields $> 10^{14}$G did not have a significant effect on the magnetorotational instability (MRI), and this MHD-induced effective viscosity was the dominant viscosity source \cite{Sun:2022vri}. 
\item Neutrino flux was able to clear out some of the baryon load in the polar regions, reducing the density by a factor of $10$ and causing the incipient jet to be launched ($\sim 15$ms) earlier compared to the neutrino-less cases
(at $\sim 25$ms after BH formation) \cite{Sun:2022vri}.
\end{enumerate}

Ciolfi et al. \cite{Ciolfi:2017uak,Ciolfi:2019fie,Ciolfi:2020hgg} have conducted simulations of NSNS mergers which result in a long-lived SMNS remnant, evolving up to 250ms post-merger. In the latest work \cite{Ciolfi:2020hgg} they use the APR4 EOS, start with initial magnetic fields of strength $10^{15}-10^{15.7}$G confined to the NS interiors, use a finest resolution of $\Delta x_{\textup{min}}= 250$m and do not include neutrino radiation transfer. The authors found that the remnant produces a collimated outflow, but strong baryon pollution in the polar regions produces a nearly isotropic density distribution of the ejecta and predicted that the terminal Lorentz factor will be far too small to correspond to a sGRB-compatible jet. They report a Lorentz factor at the edge of the simulation box ($3400$ km from the remnant) of $\sim 1.05$, and a total energy flux to rest-mass-energy flux ratio of $<10^{-2}$, which excludes the possibility of further acceleration to ultrarelativistic speeds. 

By contrast, M\"{o}sta et al. \cite{Mosta:2020hlh} carried out simulations of a NSNS  merger without magnetic fields that forms a HMNS remnant. At 17 ms after they then add a $10^{15}$ G pure poloidal magnetic field. The authors employed the LS220 EOS, with the finest resolution being $\Delta x_{\textup{min}}=250$m, and a leakage scheme for neutrino radiation. After evolving for $40-50$ ms post-merger, the authors report the formation of a mildly relativistic jet prior to collapse to a BH, and an electromagnetic luminosity of $L_{\textup{EM}} \sim 10^{51}$ erg s$^{-1}$. The authors found that neutrino cooling reduces the baryon pollution in the polar regions allowing for higher velocity outflow. Maximum Lorentz factors of $\sim 2-5$ within their simulation box of maximum extent $\sim 355$km from the remnant were observed, and taken as a conservative estimate for the asymptotic value. While this is still far below the $\Gamma \gtrsim 100$ inferred from sGRBs, it is suggested that neutrino pair-annihilation heating could reduce the baryon load and boost the Lorentz factor to the ultrarelativistic regime \cite{Fujibayashi:2017xsz,Just:2015dba}. 

Most recently Kiuchi et al. \cite{Kiuchi:2023obe} conducted a very high resolution ($\Delta x_{\textup{min}} = 12.5$m) simulation of an NSNS merger with the DD2 EOS \cite{Hempel:2009mc} resulting in a long-lived supramassive remnant.\footnote{Note that the authors of \cite{Kiuchi:2023obe} refer to a ``hypermassive" remnant, but the total mass of $2.7 M_{\odot}$ quoted is below the supramassive limit of $M_{\textup{sup}} = 2.92 M_{\odot}$ \cite{Tsokaros:2020hli} for the DD2 equation of state and a true hypermassive remnant would not survive for $> \mathcal{O}(1s)$ as claimed for this object in \cite{Kiuchi:2023obe,Fujibayashi:2020dvr}. Therefore, we suggest that the remnant in \cite{Kiuchi:2023obe} is best identified as supramassive.} Neutrino radiation was modelled using a combination of a leakage scheme and a grey M1 \cite{Shibata:2011kx} scheme \cite{Sekiguchi:2012uc,Sekiguchi:2015dma} for neutrino heating. An initial $10^{15.5}$G poloidal magnetic field is added prior to inspiral, but confined to the interior of the NSs as in \cite{Kiuchi:2015sga}. They also report the formation of a mildly relativistic jet, estimating the terminal Lorentz factor of the outflow estimated to be up to $10-20$ by the end of the simulation at $150$ms post-merger, provided the conversion of Poynting flux to kinetic energy is efficient. While the authors also report severe baryon loading, they argue that an $\alpha \Omega$ dynamo mechanism \cite{Reboul-Salze:2021rmf}, powered by the MRI, is able to amplify the large-scale magnetic field sufficiently to launch a jet. 

In this work, we further this discussion by conducting a systematic investigation of GRMHD simulations of NSNS mergers with initial magnetic fields extending from the interior into the exterior of the NSs, with a range of Arnowitt–Deser–Misner (ADM) gravitational binary masses and different spins. By fixing the EOS, we investigate binary mergers that yield remnants with mass above and below the supramassive limit, resulting either in HMNS or SMNS rapidly rotating remnants. By employing both irrotational and spinning binaries we probe the effects of spin in incipient jet launching, expanding our previous studies 
\cite{Ruiz:2019ezy} to the supramassive regime.\footnote{Note here that GRMHD studies of accretion disks around BHs (e.g.\cite{McKinney:2008ev}) suggest that a rotating BH ($a/M\gtrsim 0.4$) is a necessary, but not sufficient, condition
to produce a highly relativistic ($\Gamma \gtrsim 3$) jet. Therefore the spin of the merger remnant (either NS or BH) may be crucial to the existence or not of a jet.}

We find that our benchmark HMNS case that collapses to a BH produces an outflow consistent with a BZ incipient jet with EM luminosity $L_{\textup{EM}} \sim 10^{52}$erg s$^{-1}$ at the end of the simulation, consistent with our previous studies \cite{Ruiz:2016rai,Ruiz:2021qmm}. For the SMNS cases we also see the formation of a low-density funnel above the poles, a collimated helical magneitc field, and a mildly relativistic outflow. The SMNS cases produce an EM luminosity of $L_{\textup{EM}} \sim 10^{53}$erg s$^{-1}$ that persists for most of the SMNS cases until the end of the simulation at $\sim 50$ms post-merger, a rest-mass ejecta of $4-6\%$ of the total binary rest mass (corresponding to an estimated kilonova luminosity of $L_{\textup{knova}} \sim 10^{41}$erg s$^{-1}$). However, we also see the baryon pollution (i.e. the gas rest-mass density) is larger inside the funnel for the SMNS cases than the HMNS case after its collapse to a BH. As a result the magnetic energy per unit rest-mass-energy (the force-free parameter) is smaller. We also see indications of mixing between the low density outflow and higher density torus, leading to energy loss from the outflow. Moreover, we do not have the kind of evidence that we have for BZ-powered outflows for efficient acceleration to ultrarelativistic speeds for the SMNS central engines. We conclude that, while these outflows from SMNS central engines meet our basic criteria for \textit{incipient} jets, we cannot affirm they will produce the true ulrarelativistic jets that could give rise to a sGRB: we can only say they show \textit{jet-like structures}, and further simulations are needed on larger spatial scales to ascertain whether these can break out or remain choked due to the baryon pollution. 

We see that initial NS spin leads to a larger dynamical ejecta, a more massive and more diffuse bound torus of debris, and a larger EM luminosity compared to the irrotational binaries with the same mass. We also see that the post-merger high frequency component of the GW signal has a smaller amplitude for the spinning cases. The mass of the SMNS remnants does not appear to have a consistent effect on the outflow, although the higher mass SMNS remnants have a more compact debris torus. 

The remainder of the paper is organized as follows. In Sec.~\ref{sec:methods}, we briefly review our numerical methods and their implementation, referring the reader 
to~\cite{Etienne:2010ui,Ruiz:2021qmm,Ruiz:2020via,Ruiz:2019ezy,Ruiz:2017inq}
for further details and code tests. A detailed description of the adopted initial data and the grid structure used to evolve our NSNS systems  are given in Sec.~\ref{subsec:Initialdata} and \ref{subsec:Grid}, respectively. A suite of diagnostics used to verify the reliability of our numerical calculations is summarized in Sec.~\ref{subsec:diagnostic}. A review of our criteria for jet-sGRB compatibility is given in Sec.\ref{subsec:c_sGRB}, along with a summary of the expected magnetic amplification mechanisms appearing in NSNS mergers in sub-section~\ref{subsec:M_instabilities}.
We present our results in Sec.~\ref{Sec:Results}. Finally, we summarize our results and conclude in Sec.~\ref{sec:Conclusions}. Throughout the paper we adopt geometrized units ($G = c = 1$) except where stated explicitly. Greek indices denote all four spacetime dimensions, while Latin indices imply spatial parts only.

\section{Methods}
\label{sec:methods}
To perform the numerical simulations we use the in-house and well established \verb|Illinois GRMHD| code \cite{Etienne:2010ui,Etienne:2011re,Farris:2012ux} and the methods described in our previous works (see~e.g.~\cite{Ruiz:2021qmm,Ruiz:2020via,Ruiz:2019ezy,Ruiz:2017inq}). \verb|Illinois GRMHD| uses the \verb|Carpet| code \cite{Goodale:2003,Schnetter:2003rb} for moving-box mesh refinement. We use the Baumgarte–Shapiro–Shibata–Nakamura (BSSN) formulation of the Einstein equations \cite{Shibata:1995we,Baumgarte:1998te} with the moving-puncture gauge condition (Eqs. (2)-(4) in \cite{Etienne:2007jg}) with the damping parameter $\eta$ in the shift evolution equation set to $\eta \approx 2.0/M$, where $M$ is the total Arnowitt–Deser–Misner (ADM) mass of the system. At the boundaries we apply outgoing-wave or Sommerfeld boundary conditions to all the BSSN variables. We use fourth order centered stencils for spatial derivatives, except for shift advection terms where fourth order upwind stencils are used. Time integration is performed using the Method of Lines with a fourth order Runge-Kutta integration scheme with a Courant-
Friedrichs-Lewy (CFL) factor set to 0.45. To control spurious high frequency noise fifth order Kreiss-Oliger dissipation \cite{Baker:2006yw} is added to the evolution equations. 

For the matter evolution we evolve the equations of ideal MHD in conservative form using a high-resolution shock capturing method (see Eqs. (27)-(29) in \cite{Etienne:2010ui}) which employs the piecewise parabolic reconstruction scheme (PPM) \cite{Colella:1982ee} and the Harten, Lax, and van Leer (HLL) approximate Riemann solver \cite{Harten1983}. We evolve the magnetic field by integrating the magnetic induction equation using a vector potential in order to ensure it remains divergenceless throughout the evolution (see Eqs. (19)-(20) in \cite{Etienne:2010ui}). We also use the generalized Lorentz gauge in \cite{Farris:2012ux} to avoid the build-up of spurious magnetic fields \cite{Etienne:2011re} with a damping factor of $\xi = 2.0/M$.

\subsection{Initial data}
\label{subsec:Initialdata}

We evolve NSNSs that start from a quasiequilibrium circular orbit and then inspiral and merge. The binary consists of two identical, equal-mass NSs, constructed using the Compact Object CALculator (\verb |COCAL|) code (see e.g.~\cite{Tsokaros:2015fea,Tsokaros:2018dqs}) with a soft Skyrme-Lyon (SLy) equation of state (EOS) \cite{Douchin:2001sv} modeled using a piecewise polytropic representation as in \cite{Read:2008iy}. This EOS is a largely arbitrary choice chosen to be consistent with our previous works. Nonetheless, SLy remains a realistic EOS candidate broadly consistent with observational constraints as discussed in \cite{Ruiz:2021qmm}, Sec. II. B. The predicted maximum gravitational mass for an isolated, cold, spherical NS is $M_{\textup{TOV}} = 2.049 M_{\odot}$ for SLy, consistent with both the $M_{\textup{TOV}}$ limits from the observations of pulsars PSR J0740+6620 \cite{NANOGrav:2019jur,Riley:2021pdl}, PSR J1614-223 \cite{Fonseca:2016tux}, PSR J0348+0432 \cite{Antoniadis:2013pzd} and constraints from GW170817. The predicted radius of cold, spherical neutron star of typical mass $1.4 M_{\odot}$ with SLy is 11.46 km, consistent with constraints on the radius of PSR J0740+6620 \cite{Pang:2021jta} and inferred constraints on the progenitors of GW170817 \cite{LIGOScientific:2018cki}. The low estimated tidal deformability of a $1.4 M_{\odot}$ mass NS inferred from GW170817 by a LIGO/Virgo analysis \cite{LIGOScientific:2018cki} also favours a soft EOS like SLy over stiff alternatives. However, the large radius of PSR J0030+0451 inferred by a NICER analysis \cite{Miller:2019cac,Riley:2019yda} instead suggests at stiffer EOS in tension with SLy, as does the high NS mass of $2.59^{+0.08}_{-0.09}$ inferred for the secondary object in GW190814 \cite{LIGOScientific:2020zkf} if that object was indeed a NS at merger \cite{Tsokaros:2020hli}, and EOS constraints remain a matter of debate. The full list of critical masses $M_{\textup{TOV}},M_{\textup{sup}},M_{\textup{thresh}}$, in terms of both rest mass and ADM gravitational mass, for a cold SLy are given in Table \ref{tab:SLy_mass_lims}.

To account for shock heating during the merger we also add a thermal component to the EOS on top of the cold SLy, as described in \cite{Ruiz:2021qmm} Eqs (1)-(3). We write the total pressure as $P = P_{\textup{th}} + P_{\textup{cold}}$ where $P_{\textup{cold}} = P_{\textup{SLy}}(\rho_0)$ is the cold SLy component and $P_{\textup{th}}$ is a thermal component given by 
\begin{equation}
    {P_{\textup{th}} = (\Gamma_{\textup{th}}-1) \rho_0(\epsilon - \epsilon_{\textup{cold}}),}
\end{equation}
where $\epsilon_{\textup{cold}}$ is the internal energy calculated from the SLy EOS and $\Gamma_{\textup{th}} = 5/3$ appropriate for ideal nonrelativistic baryons \cite{Bauswein:2010dn,Paschalidis:2011ez}, as in \cite{Ruiz:2021qmm}. 

\bgroup
\def\arraystretch{1.0}%
\begin{table}
\begin{tabular}{l|cc}
    & $\;$rest mass $M_0$$\;$$\;$  &  $\;$$\;$ADM mass $M$ \\
    & [$M_{\odot}$] & [$M_{\odot}$] \\[.05cm] 
    \hline
    $M_{\textup{TOV}}$ & 2.46 & 2.06 \\
    $M_{\textup{sup}}$ & 2.96 & 2.49 \\
    $M_{\textup{thresh}}$ & $\dots$ & $\sim 2.8$
\end{tabular}
\caption{Critical mass limits for the SLy EOS \cite{Read:2008iy,Tsokaros:2020hli,Kolsch:2021lub,Koppel:2019pys}. The estimate for the gravitational $M_{\textup{thresh}}$ is from a series of GRHD merger simulations of initially irrotational neutron stars (see \cite{Kolsch:2021lub} for details).}
\label{tab:SLy_mass_lims}
\end{table}
\egroup

To explore how the mass and spin of the system, and therefore the properties and nature of the remnant, affect jet formation we consider cases with five different initial binary ADM masses ($M$) from $2.40 M_{\odot}$ to $2.70 M_{\odot}$ in Table \ref{tab:initial_NS}. For each of the bottom four masses we also explore the effect of the NS spin, evolving one case where the stars are irrotational (denoted IR) and one where they are spinning (denoted SP) with dimensionless spins $\chi := J_{\rm ql}/(M/2)^2 \approx 0.27$, where $J_{\rm ql}$ is the quasilocal angular momentum of the NS \cite{Tsokaros:2018dqs}. Note that this formula is only strictly valid for widely separated NSs where the gravitational potential energy interaction energy is negligible. Observations of binary pulsar systems suggest there are at least some NSNSs where the stars have non-zero spins at merger \cite{Zhu:2017znf}. While the spins we use here are significantly larger than those inferred from such observed binaries (which are fewer than twenty), they provide a proof-of-principle study of the impact of NS spin. 
Full details are given in Table \ref{tab:initial_NS}. 
\begin{figure}
    \centering
    \includegraphics[width=0.45\textwidth]{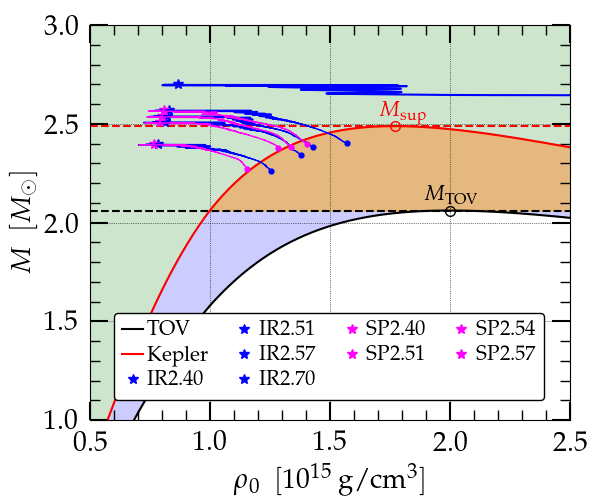}
    \caption{Mass vs rest-mass density for isolated NSs with the SLy EOS. The black and red curves show
    the TOV and mass-shedding limit of spherical and uniformly rotating
    NSs, respectively. Blue (magenta) stars correspond to our initial irrotational (spinning) binaries, where 
    $M$ denotes the approximate total gravitational mass of the NS binary or its resulting merger remnant and $\rho_0$ the maximum density in each NS companion. Blue (magenta) dots
    correspond to the end state of the corresponding system, with solid blue (magenta) lines depicting the evolution
    trajectories of the systems, through merger up until the end of our simulations.
    }
    \label{fig:Mrho}
\end{figure}
Our models are shown in a mass vs rest-mass density diagram
in Fig. \ref{fig:Mrho} where now $M$ is the approximate gravitational mass of the NS binary or, following merger, the resulting remnant and $\rho_0$ the maximum density of each NS.
Blue (magenta) stars correspond to the initial irrotational (spinning) binaries, while blue (magenta) dots correspond to the final remnant object (everything that exists in the computational domain). Shaded regions correspond to normal spinning NSs (violet), SMNSs (orange), and stars supported by differential rotation (green). The region above the red dashed line denoting $M = M_{\textup{sup}}$ corresponds to hypermassive stars. 
Solid blue (magenta) lines depict the evolution of the gravitational mass $M$ and maximum rest-mass density $\rho_0$ across merger all the way to the end of our simulations. In interpreting this diagram we should keep in mind i) the mass is computed via a surface integral at a finite radius, not at spatial infinity, and ii) this mass ($M$) includes not only the rotating NS remnant, but also the inner portion of the disk around it. Having said that, the mass of the rotating remnant NS (without the disk) should be somewhere lower than the endpoints which are denoted with blue or magenta dots, and therefore they all lie in the supramassive regime, except for the highest mass model (IR2.70) which leads to a HMNS and finally a BH. In addition, the merger remnant has a finite temperature, and therefore the (cold) mass limits in Fig. \ref{fig:Mrho} underestimate the dynamical ones.

\begin{table}
\begin{ruledtabular}
\begin{tabular}{c|ccccc}
Case & $M$[$M_{\odot}$] & $M_{\textup{NS}}$[$M_{\odot}$] & $R_x$[km]  & $\chi$ & ${\Omega}M$ \\
\colrule
IR2.40 & 2.40 & 1.33 & 9.39  & 0.00 & 0.026 \\
IR2.51 & 2.51 & 1.39 & 9.28  & 0.00 & 0.028 \\
IR2.54 & 2.54 & 1.41 & 9.25  & 0.00 & 0.029 \\
IR2.57 & 2.57 & 1.43 & 9.21  & 0.00 & 0.029 \\
IR2.70 & 2.70 & 1.51 & 9.05  & 0.00 & 0.030 \\
\colrule
SP2.40 & 2.40 & 1.32 & 9.70  & 0.27    & 0.025 \\
SP2.51 & 2.51 & 1.39 & 9.57  & 0.27    & 0.027 \\
SP2.54 & 2.54 & 1.41 & 9.54  & 0.27    & 0.028 \\
SP2.57 & 2.57 & 1.43 & 9.50  & 0.26    & 0.028
\end{tabular}
\end{ruledtabular}
\caption{Summary of the initial properties of the NSNS cases. We list the name of the case, the asymptotic gravitational (ADM) mass of the binary system $M$, the rest mass of each star $M_{\textup{NS}}$, the equatorial coordinate radius of each star meaasured along the axis of the binary $R_x$, the dimensionless NS spin $\chi$, and the dimensionless quantity $\Omega M$ where $\Omega$ is the orbital angular velocity of the binary. The initial coordinate separation is set to $3.98 R_x$ for the cases with mass $2.40 M_{\odot}$ to $2.57 M_{\odot}$ and $4.22 R_x$ for the $2.70 M_{\odot}$ mass case.}
\label{tab:initial_NS}
\end{table} 

\begin{figure*}
\begin{tabular}{cc}
  \includegraphics[width=0.5\textwidth]{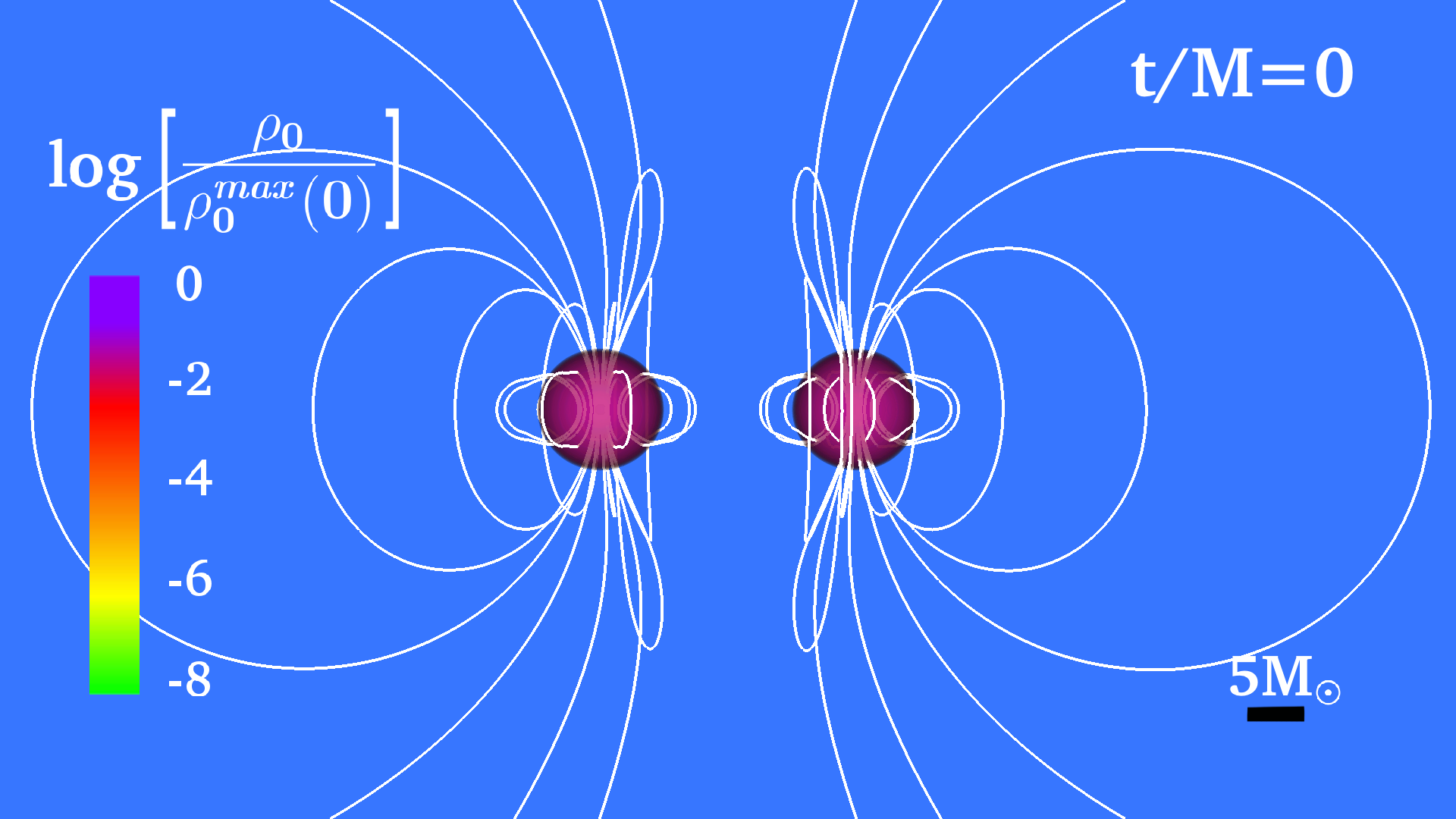} &   \includegraphics[width=0.5\textwidth]{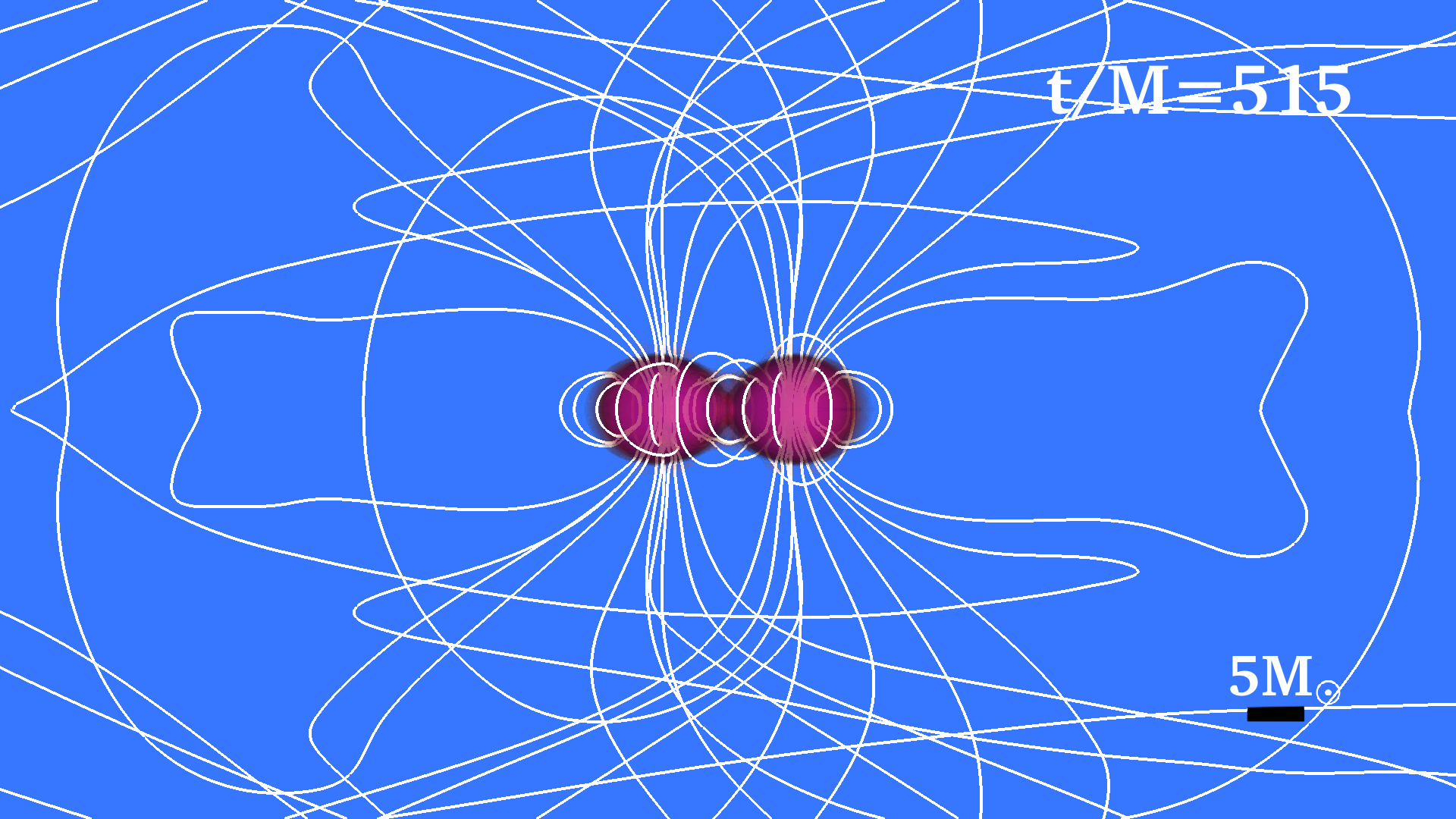} \\
 \includegraphics[width=0.5\textwidth]{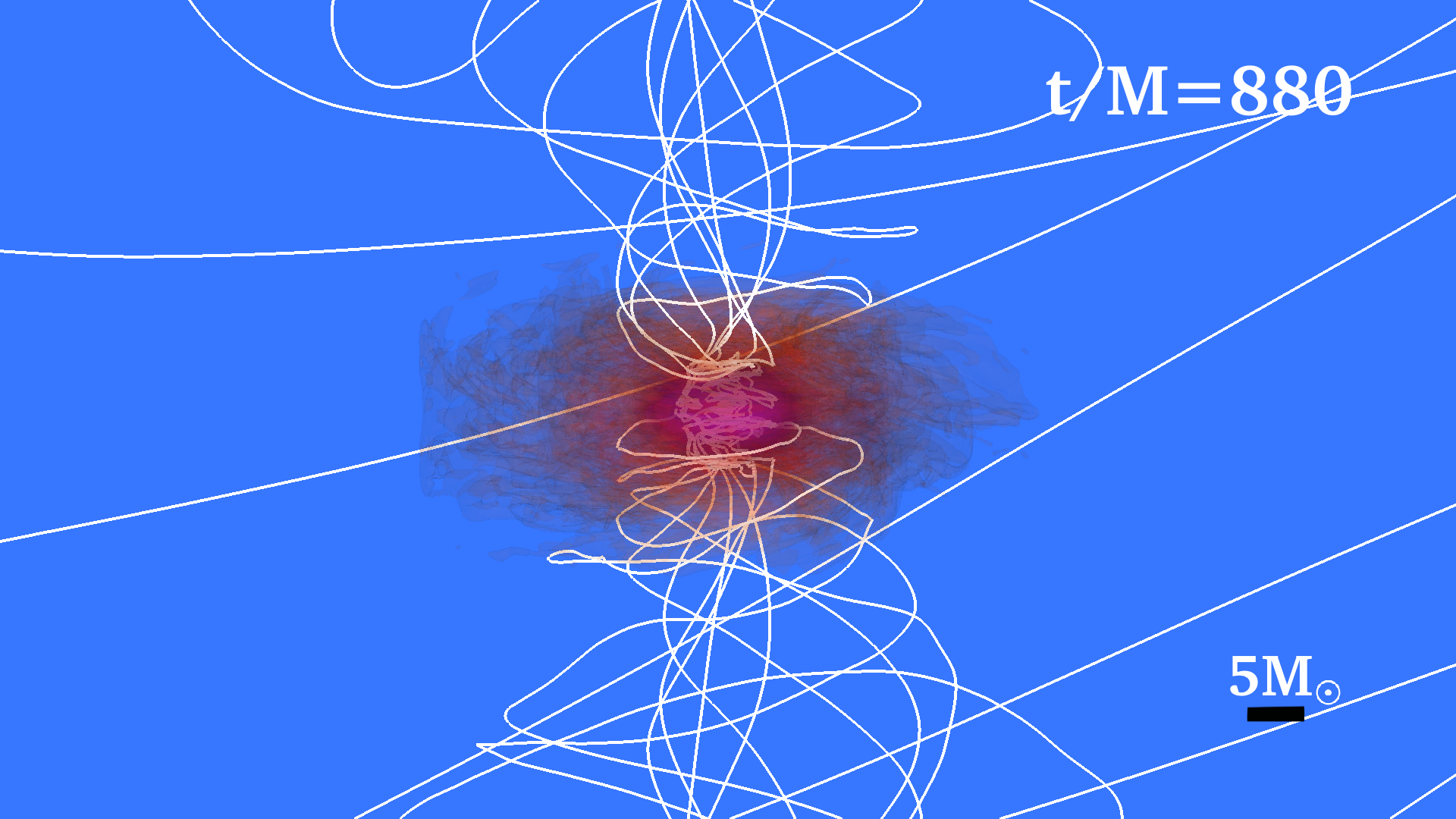} &   \includegraphics[width=0.5\textwidth]{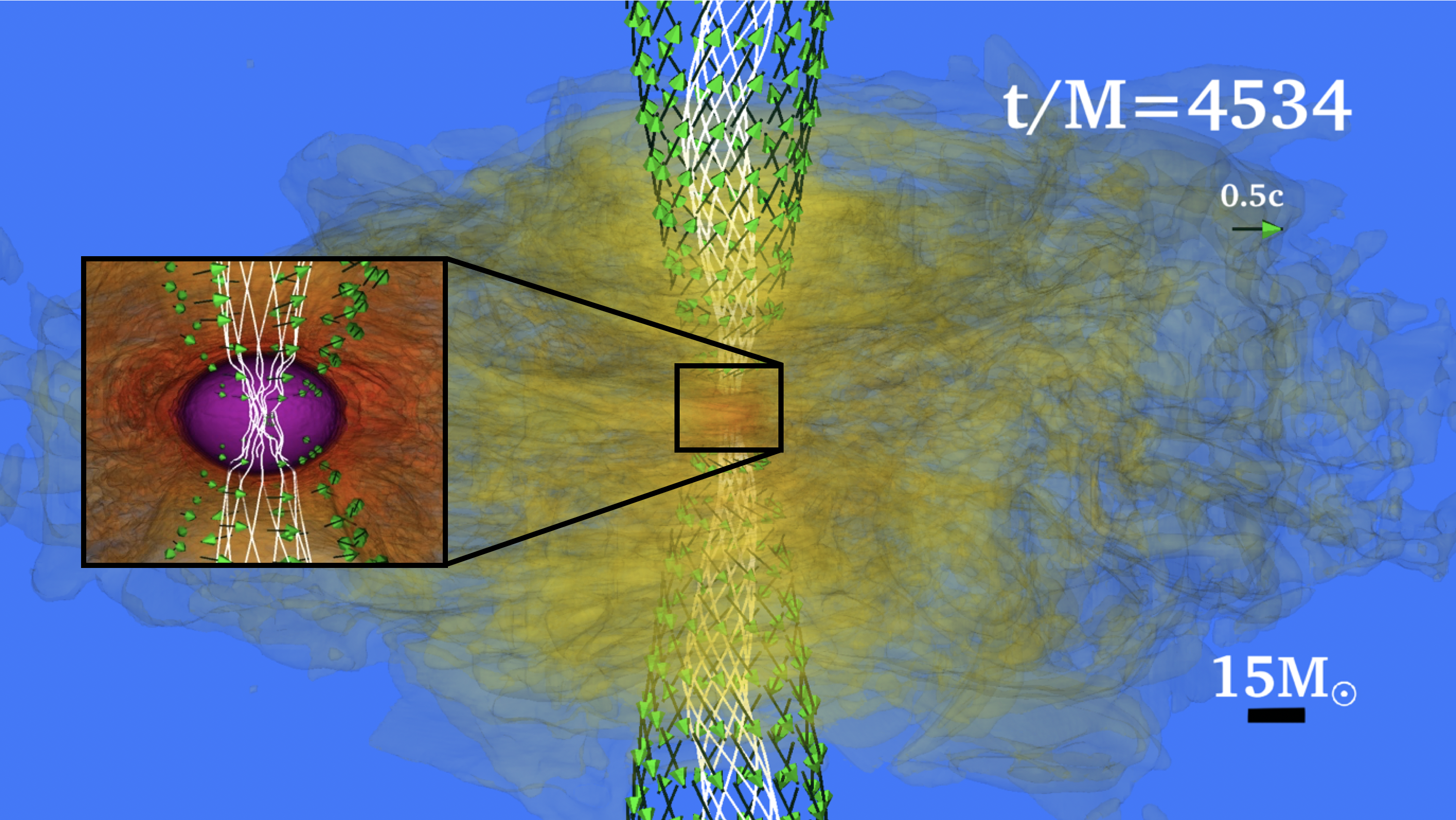} 
\end{tabular}
\caption{3D snapshots of the evolution of the rest-mass density for the IR2.54 case (see Sec.~\ref{subsec:Initialdata}) at four instances in time. The images in the upper left, upper right and lower left show the baryonic rest-mass density (for $\rho_0/\rho^{\textup{max}}_0(t=0)\geq 10^{-2}$) and the magnetic field lines at the start of the simulation, when the stars make contact, and shortly after the merger respectively. The lower right image shows the torus of matter that forms around the central SMNS remnant at $\sim 50$ms near the end of the simulation. The yellow torus shows the region with $\rho_0/\rho^{\textup{max}}_0(t=0)\gtrsim 10^{-6}$, the white lines show the helical collimated field lines emerging from the poles, and the green arrows showing the fluid velocity along the evacuated magnetically dominated funnel. We also show an insert with the front half cut away on the meridional plane, showing the remnant itself in purple and the higher density parts of the torus in red-orange. Here $M =2.54 M_{\odot}$ and $M_{\odot} = 4.9\times 10^{-3}\textup{ms} = 1.4$km.
}
\label{fig:3D_plots}
\end{figure*}

Each NS is initially endowed with a poloidal pulsar-like magnetic field, following \cite{Farris:2012ux,Paschalidis:2013jsa,Ruiz:2017inq}, generated by the magnetic vector potential 
\begin{equation}
    A_{\phi} = \frac{15\pi \varpi^2 I_0 r^2_0}{23(r^2_0 + r^2)^{3/2}}\left[1 + \frac{15r^2_0(r^2_0+\varpi^2)}{8(r^2_0+r^2)^2}\right],
\end{equation}
which approximately corresponds to that generated by a current loop inside the NS with radius $r_0$ and current $I_0$ (Fig. \ref{fig:3D_plots} upper left). Here $r^2 = \varpi^2 + z^2$ and $\varpi^2 = (x - x_{\rm NS})^2 + (y - y_{\rm NS})^2$, where the center of the star $(x_{\rm NS},y_{\rm NS},0)$ is determined by the coordinate of maximum rest-mass density. We set $I_0$ and $r_0$ such that magnetic field as measured by a normal observer at the NS pole is $B_{\textup{pole}} = 10^{15.2}$G and the maximum value is $B_{\textup{max}} \sim 10^{16.7}$G in the NS center. The maximum value of the magnetic-to-gas-pressure ratio in the NS interior is $\beta^{-1} := P_B/P_{\textup{gas}} = 0.0023$. The magnetic field at the pole is significantly larger than the surface magnetic field strengths of $10^8 - 10^{12.2}$G expected for NSs in binary systems as inferred from observations of binary pulsars \cite{Tauris:2017omb,Lorimer:2008se}. However, this value is chosen to model the field strengths expected due to the exponential growth from magnetic instabilities, initially driven by the Kelvin-Helmholtz Instability (KHI) at the shearing surface when the stars first collide. This instability boosts the magnetic energy by almost a factor of 10 within several ms, until the instability saturates or the shear surface is destroyed by shocks \cite{Kiuchi:2015sga}. The magnetic energy growth rate in the linear regime is inversely proportional to the minimum resolvable wavelength, and thus to the numerical resolution, at least down to a resolution $\Delta x_{\textup{min}} \sim 12.5$m \cite{Price:2006fi,Kiuchi:2023obe}. The magnetorotational instability (MRI), magnetic winding and potentially the Rayleigh-Taylor instability also work to boost the magnetic field. Detailed special relativistic and approximate-GR simulations have shown that the field can be amplified to magnetar levels of $\gtrsim 10^{15}$G as the stretching and folding of the magnetic field lines converts kinetic energy to magnetic \cite{Price:2006fi,Rosswog:2006ue,Obergaulinger:2010gf,Oechslin:2006uk,Zrake:2013mra}. Very high resolution GRMHD simulations have shown similar results, with the $\Delta x_{\textup{min}} = 37$m simulation of Aguilera-Miret et al. \cite{Aguilera-Miret:2020dhz} showing amplification from $10^{11}$ to $10^{16}$G within the first 5ms post-merger, and the $\Delta x_{\textup{min}} = 17.5$m simulation of Kiuchi et al. (2015) \cite{Kiuchi:2015sga} where an initial field strength of $10^{13}$G is amplified to create a large-scale $\gtrsim 10^{16}$G poloidal field due to a combination of the KHI and the MRI. Computational cost limits the resolution we can use here, so the KHI growth rate in our work is consequently suppressed compared to expected physical reality. Following our previous works \cite{Ruiz:2016rai,Ruiz:2017inq,Ruiz:2021qmm} we therefore adopt the artificially strong initial magnetic field to compensate, as is common practice in GRMHD simulations of NSNS mergers~\cite{Ciolfi:2019fie}. Notice that this magnetic field modifies the pressure of the system in $<1\%$ and hence it does not have a significant impact on the NS structure or the inspiral phase of the binary. During this phase the magnetic field is simply advected by the fluid. While it has been suggested that purely poloidal fields are unstable on an Alf\'{e}n time scale \cite{Ciolfi:2011xa,Lasky:2012ju}, our previous works have shown that a strong, large-scale poloidal field is a requirement for jet launching (e.g. \cite{Etienne:2012te,Ruiz:2020via}), and this idealized topology both allows us to resolve the MRI instability (as the MRI wavelength is proportional to $\vert b^P \vert$) and serves as a useful model. Hence, while the true pre-merger magnetic field structure inside neutron stars remains uncertain \cite{Braithwaite:2005md,Bilous:2019knh,Tsokaros:2021pkh} a strong poloidal field provides a ``best-case" scenario for jet launching. 

An alternative approach using large-eddy simulations and subgrid-scale models is discussed in \cite{Giacomazzo:2014qba,Aguilera-Miret:2020dhz,Palenzuela:2021gdo,Aguilera-Miret:2021fre,Aguilera-Miret:2023qih,Izquierdo:2024rbb}. The idea is to compensate for the limited resolution by including additional terms into the evolution equations that attempt to capture the otherwise unresolved subgrid dynamics. The advantage is that it allows you to start with smaller magnetic fields, closer to those of known pulsars, and it is suggested that it is better able to model the small scale quasi-isotropic turbulent field generated via the KHI \cite{Aguilera-Miret:2020dhz}, and that it is preferable to using artificially strong initial poloidal fields which the authors in \cite{Aguilera-Miret:2023qih} argue could lead to unrealistic outcomes. However, the downside is that the results may depend heavily on the choice of subgrid model and the coefficients chosen for the subgrid terms, making it unclear whether such models accurately capture the true physics. 

As in our previous studies \cite{Paschalidis:2014qra,Ruiz:2016rai,Ruiz:2019ezy,Ruiz:2021qmm}, to reliably evolve the exterior magnetic fields within the assumptions of ideal MHD we initially add a low-density artificial atmosphere exterior to the NSs in regions where the magnetic field dominates over the fluid pressure gradient. The density of this artificial atmosphere is chosen such that at $t=0$ the plasma parameter $\beta$ satisfies $\beta = P_{\textup{gas}}/P_{\textup{magnetic}} = 0.01$, with an additional density floor of $\rho_0^{\textup{min}} = 10^{-10}\rho_0^{\textup{max}}$, where $\rho^{\textup{max}}_0$ is the maximum value of the initial rest-mass density of the system. Further implementation details can be found in \cite{Ruiz:2018wah} Sec. II. B. The artificial atmosphere increases the total rest mass of the system by $\lesssim 2\%$, and was shown previously to have a negligible effect on the dynamical evolution \cite{Paschalidis:2014qra}.

\subsection{Grid structure.}
\label{subsec:Grid}
The grid structure uses the ``moving-box" approach, with two sets of nested grids centered on each star. There are nine refinement levels of nested grids differing in size and resolution by a factor of two, plus the coarsest level which covers the whole simulation box. The simulation box is a half-cube (using equatorial symmetry across the $xy$ plane) of spatial extent $L_0/2 \simeq 5748\textup{km} \simeq 3891 M_{\odot}$, where $L_0$ is the total width, and grid spacing 
$\Delta x_0 \simeq 46\textup{km} \simeq 31 M_{\odot}$, so that each subsequent level has half-width 
$L_n/2 \simeq 5748/2^n \textup{km}$
and grid spacing  
$\Delta x_n \simeq 46/2^n \textup{km}$
for $n = 1,2 \dots 9$. The maximum resolution is 
$\Delta x_9 = \Delta x_{\textup{min}} \simeq 90\textup{m}$. The number of grid points covering the equatorial diameter of the NS, denoted $N_{NS}$, is then between $200$ and $214$ for the most and least compact cases, respectively. We use the same grid configuration for all the cases. Note that the resolution used here is a factor of $\sim 1.25$ finer than that used for SLy models in \cite{Ruiz:2021qmm}. When two grid boxes overlap they are replaced by a combined box centered on the center of mass of the system.  

\subsection{Diagnostics}
\label{subsec:diagnostic}
 
The $M = 2.70 M_{\odot}$ case forms a HMNS remnant, which is the only one which collapses to a BH before our simulation terminates. After collapse, we use the \verb|AHFinderDirect| thorn \cite{Thornburg:2003sf} to track the apparent horizon and estimate the BH mass and dimensionless spin using the formalism of \cite{Dreyer:2002mx}. 

We extract the GW signal by computing the Weyl scalar $\Psi_4$ using the \verb | Psikadelia | thorn, then decompose it into $s = -2$ spin-weighted spherical harmonic modes extracted over spherical surfaces at seven different extraction radii between $120M_{\odot}$ and $840M_{\odot}$. We then convert these values to $h_{+/\times}$ strain polarizations and compute the energy and angular momentum flux radiated away in GWs (for further details see \cite{Ruiz:2007yx}). The GW luminosity can be obtained from the $\Psi_4$ Weyl scalar as 
\begin{equation}
    L_{\textup{GW}} = \lim_{r\rightarrow\infty}\frac{r^2}{16\pi}\int \left\vert\int^t_{-\infty}\Psi_4 \dd t'\right\vert^2 \dd \Omega, 
\end{equation}
which we approximate via a surface integral at a finite radius in the wave zone.

We monitor the outflow of matter by computing the unbound rest mass outside a radius $r_0$ respectively as 
\begin{equation}
\begin{split}
    M_{\textup{esc}} =& \int_{r>r_0} \rho_{*} \Theta(-u_t-1)\Theta(v^r)\dd x^3 \\
    &+ \int^t_{t'=0}\int_{\delta\mathcal{D}}\rho_{*} \Theta(-u_t-1)\Theta(v^r)v^i\dd S_i \dd t',
    \label{eq:esc_mass}
\end{split}
\end{equation}
where $\rho_* = \sqrt{-g}\rho_0 u^t$, $\dd S_i$ is the surface element on the sphere, and the Heaviside functions $\Theta$ ensure we only include material with a positive specific energy $E = -u_t - 1$ (i.e. unbound material) with a positive radial velocity. Note that we also add in the contribution from the rest mass leaving the boundary of the simulation domain, $\delta \mathcal{D}$, although this contribution is $\lesssim 10^{-5} M_{\odot}$ by the end of the simulation. Here $\rho_0$ is the rest-mass density, $g$ the determinant of the 4-metric, $u^{\mu}$ the four-velocity of the fluid and $v^r$ the radial component of the three-velocity. We examine radii $r_0 = 30M,50M,70M,100M$ and confirm that the difference between them is less than $\lesssim 2.5\%$. 

\begin{figure}
    \centering\includegraphics{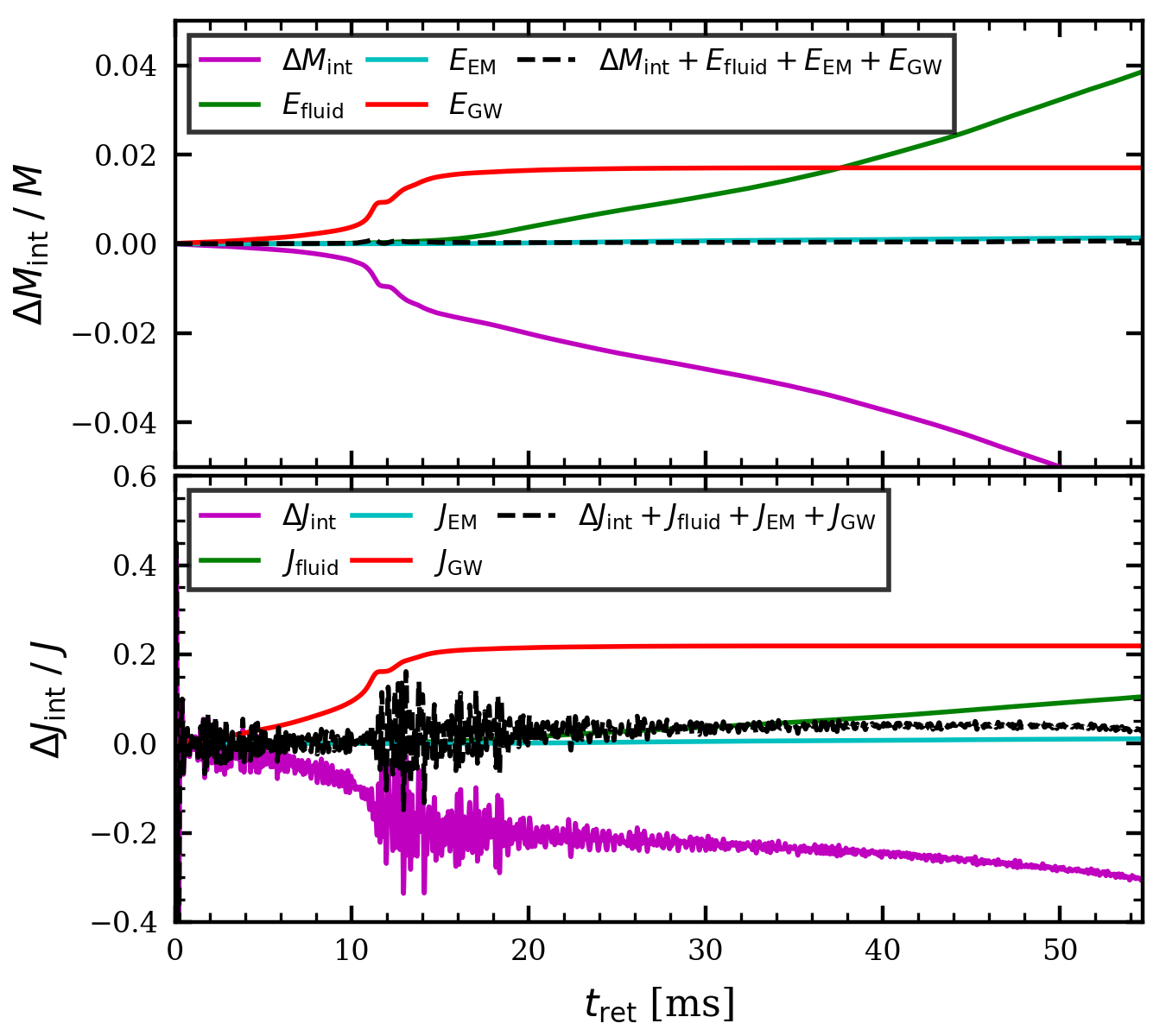}
    \caption{Conservation of the mass and angular momentum integrals $M_{\textup{int}}$ and $J_{\textup{int}}$ for the representative SP2.54 case vs, retarded time $t_{\textup{ret}} := t - r_*$, where $r_*$ is the extraction radius in tortoise coordinates $r_* = r + 2M\ln(r/(2M)-1)$. The magenta lines denote the change in the surface integrals Eqs. (37) and (39) in \cite{Etienne:2011re} evaluated on spherical surfaces at $r_{\textup{ext}}\simeq 273M$. The cyan line denotes the total EM energy (upper plot) and $z$ component of the total EM angular momentum (lower plot) that has passed across that surface. The green and red lines do the same for the fluid and GWs respectively. The black dashed line gives the total energy in the top panel, and total angular momentum in the bottom panel that should be conserved, adding in the lost energy and angular momentum.}
    \label{fig:MJ_conservation}
\end{figure}

In addition, we compute the rest-mass outflow 
\begin{equation}
\dot{M}_0 = \int \rho_* v^{i} \dd S_i \ ,    
\label{eq:rmo}
\end{equation}
the fluid energy luminosity \cite{Gottlieb:2023est}
\begin{equation}
L_{\textup{fluid}} = \int \sqrt{-g}(-T^{i\;(\textup{fluid})}_t - \rho_0 u^i)\dd S_i  \ ,
\label{eq:fel}
\end{equation}
(where in Eq. \eqref{eq:fel} we subtract the contribution from the rest-mass-energy flux), and the EM Poynting luminosity 
\begin{equation}
L_{\textup{EM}} = - \int \sqrt{-g}T^{i\;(\textup{EM})}_t \dd S_i\ , 
\label{eq:pl}
\end{equation}
over ten spherical surfaces with radii equally spaced from $\simeq 57$km to $\simeq 4320$km, or approximately $15M$ to $1200M$, as well as the corresponding fluxes over the outer surface of the simulation domain (note that we do not calculate the thermal emission from the gaseous debris). We monitor the conservation of the total mass and total angular momentum $M_{\textup{int}}$ and $J_{\textup{int}}$ (defined via Eqs. (37) and (39) in \cite{Etienne:2007jg} with an integral over a spherical surface of finite radius) which correspond to the ADM mass and the $z$ component of the ADM angular momentum respectively when evaluated at spatial infinity ($r=\infty$). Consistent with \cite{Tsokaros:2019anx,Ruiz:2019ezy} we find that mass is conserved to $< 1\%$ and angular momentum is conserved to $\lesssim 5\%$, as shown in Fig. \ref{fig:MJ_conservation}. 
We also monitor the total magnetic energy $E_{\rm mag} = \int n^{\mu} n^{\nu}T^{\textup{EM}}_{\mu\nu} \dd V$ outside the BH horizon (if present) measured by a normal observer and its growth over time, as well as the effective Shakura–Sunyaev parameter \cite{Shakura:1973boa} for the effective viscosity due to the magnetic field 
\begin{equation}
\alpha_{\textup{SS}} \sim \frac{\textup{magnetic stress}}{\textup{pressure}} =
\frac{T^{\textup{EM}}_{\hat{r}\hat{\phi}}}{P}
\label{eq:alphass}  
\end{equation}
where $T^{EM}_{\hat{r}\hat{\phi}} = e_{\hat{r}}^{\ \mu}e_{\hat{\phi}}^{\ \nu}T_{\mu\nu}^{\textup{EM}}$ is the $r\phi$ component of the electromagnetic stress-energy tensor in the local comoving frame and $e_{\hat{i}}^{\ \mu}$ the corresponding basis of local tetrads (see Eq. (26) in \cite{Penna:2010hu}). 

\subsection{Criteria for a sGRB-compatible jet}
\label{subsec:c_sGRB}

As discussed in Sec.\ref{sec:Intro}, to be consistent with sGRB observations, the central engine needs to produce a an ultrarelativistic collimated jet with a maximum Lorentz factor of $\Gamma \gtrsim 20$ and, to match the kilonova observations, an ejecta mass of $\gtrsim 1\%$ of the total NSNS mass~\cite{Coulter:2017wya,Shappee:2017zly,Arcavi:2017xiz}. One highly non-trivial challenge in analyzing the result of GRMHD merger simulations is estimating the terminal Lorentz factor $\Gamma_{\infty}$, as our simulations have a finite spatial and temporal extent while we expect the velocity to increase with distance from the central engine (for instance, for a simple model of an ideal force-free paraboloidal jet the maximum Lorentz factor goes with distance along the jet $z$ as $\Gamma_{\textup{max}} \sim 0.3 (z/M)^{1/2}$ \cite{Tchekhovskoy:2008gq,McKinney:2008ev,Narayan:2010bi}). The physics of such jets, whether in the context of sGRB progenitors or active galactic nuclei, has been explored extensively through both numerical MHD and GRMHD studies (see e.g. \cite{Tchekhovskoy:2008gq,McKinney:2008ev,Narayan:2010bi,Fernandez:2018kax,Nathanail:2020hkx,Gottlieb:2022sis,Mizuno:2022vqa,Perucho:2023bri} and the references cited therein) and semi-analytic studies (see e.g. \cite{Vlahakis:2003si,Beskin:2005uq,Komissarov:2010zm,Lyubarsky:2010yd}) which make use of the assumptions of ideal MHD, special relativity, axisymmetry, steady state and radial self-similarity. For a steady state, axisymmetric ideal MHD flow integrating the equations of motion results in a number of conserved quantities along the poloidal component $\boldsymbol{B}_p$ of the magnetic field lines, which are parallel to the poloidal component of the fluid velocity $\boldsymbol{v}_p$ \cite{Vlahakis:2003si,Narayan:2010bi,Lyubarsky:2010yd}. One of these quantities is the ratio of total energy flux to rest-mass flux along a bundle of field lines, given by 
\begin{align}
    \mu :=& \frac{\textup{energy flux}}{\textup{rest-mass flux}}= \;\frac{\Gamma^2 \rho_0 h v_p + \tfrac{1}{4\pi}( \boldsymbol{E} \times \boldsymbol{B})\cdot \hat{\boldsymbol{v}}_p}{\Gamma \rho_0 v_p},
    \nonumber\\
    =& \;\Gamma + \Gamma(h - 1) + \sigma \Gamma, 
\end{align}
where in the second line we have decomposed it into the contributions from the specific kinetic $+$ rest-mass energy $\Gamma$, the enthalpy contribution $\Gamma(h-1)$, and the EM Poynting flux contribution $\sigma \Gamma = \tfrac{1}{4\pi}\vert \boldsymbol{E} \times \boldsymbol{B}_{\phi}\vert /(\Gamma \rho_0 v_p)$, where $\sigma$ is the magnetization parameter \cite{Narayan:2010bi,Lyubarsky:2010yd}. The gravitational energy is typically neglected, assuming a flat Minkowski metric. At the base of the jet the flow is sub-relativistic with $\Gamma \approx 1$, with the energy flow dominated by the EM Poynting flux with $\sigma \gg 1$ and $\mu \sim \sigma$ (for $\mu \gg 1$ the enthaply component is a sub-dominant contribution). As the fluid is accelerated upwards along the magnetically dominated low density funnel, magnetic energy is converted to kinetic energy, so $\sigma$ decreases as $\Gamma$ increases in such a way as to keep $\mu$ constant. If the acceleration were perfectly efficient the final asymptotic Lorentz factor would be $\Gamma_{\infty} \approx \mu$. However, in reality it is likely that not all of the energy will be converted, and $\Gamma_{\infty} < \mu$. Calculations using the self-similar model predict that the final asymptotic state is cylindrical flow parallel to the jet axis \cite{Vlahakis:2003si} with $\Gamma_{\infty} \approx \mu/2$ and rough equipartition between the kinetic and EM energy. 

We can express $\mu$ and $\sigma$ in terms of the ratio of the EM energy density $\rho_{\rm B} = b^2/2 = B^2_{\textup{co}}/8\pi$ (where $b^{\mu} := B^{\mu}_{\textup{co}}/\sqrt{4\pi}$ for comoving magnetic field $B^{\mu}_{\textup{co}}$) to the rest-mass density $\rho_0$. For a strongly poloidal flow with $v_p \gg v_{\phi}$ we have 
\begin{equation}
    \sigma \approx \frac{\sin^2 \zeta}{\Gamma^2(1 - v^2 \sin^2 \zeta)}\frac{b^2}{\rho_0},
\end{equation}
where $\zeta$ is the angle between the magnetic field and the fluid velocity. The strongest acceleration occurs where the magnetic field is tightly coiled and $B_{\phi} \gg B_{p}$ with $\sin\zeta \approx 1$, so for these regions near the base of the outflow where $\Gamma \sim h \sim 1$ we can estimate 
\begin{equation}
    \Gamma_{\infty} \approx \;\frac{\mu}{2} \approx \frac{\sigma}{2} \approx \frac{b^2}{2\rho_0} = \frac{\rho_B}{\rho_0},
\end{equation}
for $b^2/(2\rho_0) \gg 1$ using the equipartition result. Based on this, as in our previous works \cite{Paschalidis:2014qra,Ruiz:2021qmm}, we define an \textit{incipient} jet as a tightly collimated, mildly relativistic outflow which is driven by a tightly wound ($\sin \zeta \approx 1$), helical, force-free ($b^2/(2\rho_0) \gg 1$)
magnetic field. There remains the issue of where we measure $b^2/(2\rho_0)$, as near the central engine the assumption of a flat spacetime breaks down. Kiuchi et al. \cite{Kiuchi:2023obe} and Metzger et al. \cite{Metzger:2007cd} consider the magnetization $\sigma$ evaluated at the light cylinder radius $R_{\textup{LC}} \equiv c/\Omega$. They assume perfect acceleration efficiency, with all the magnetic energy converted to magnetic, so estimate $\sigma_{\textup{LC}} = \sigma(R=R_{\textup{LC}}) \approx \Gamma_{\infty}$ with no factor of $1/2$. 
\begin{figure}
\begin{tabular}{c}
  \includegraphics[width=0.5\textwidth]{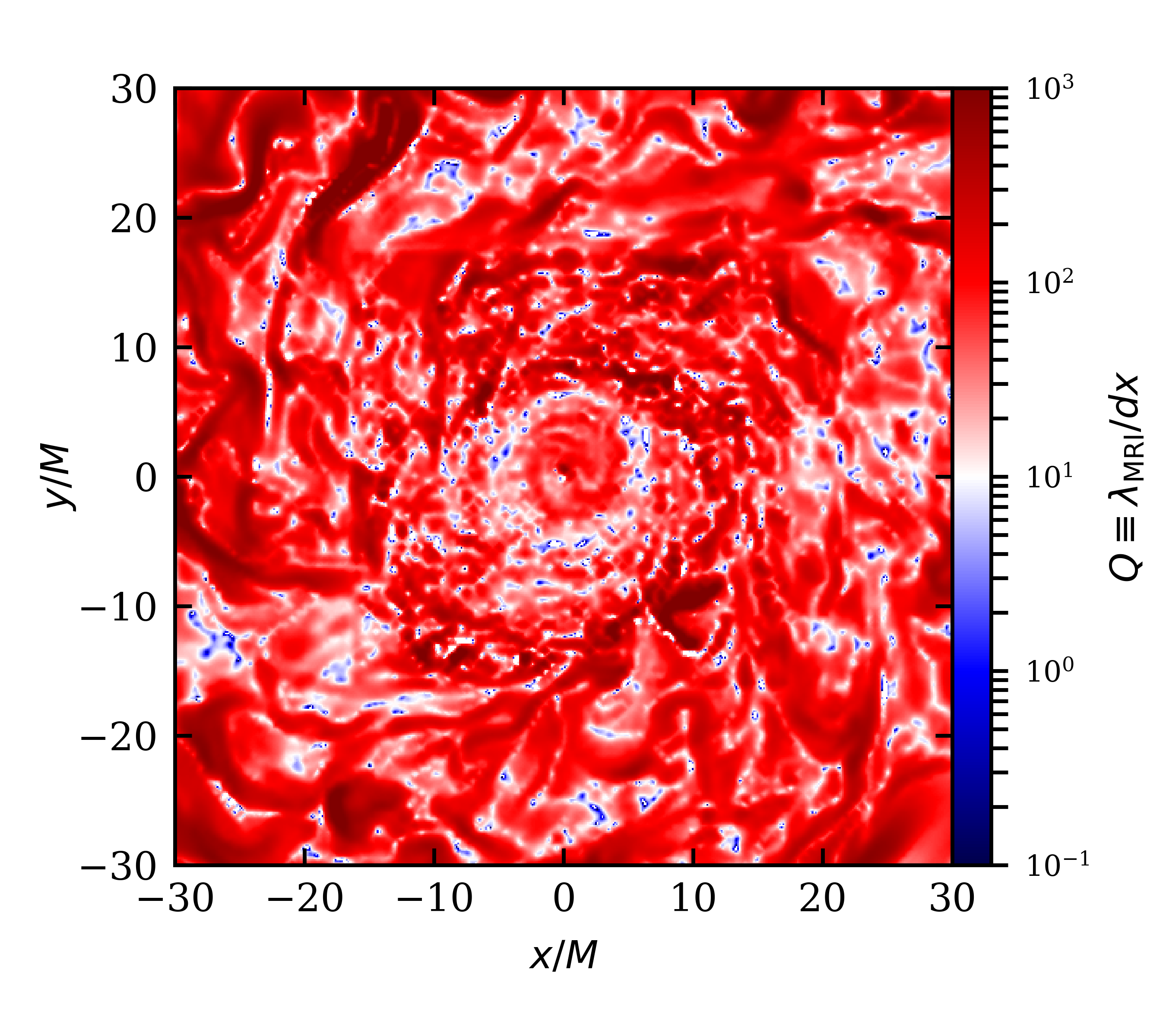} \\
  \includegraphics[width=0.5\textwidth]{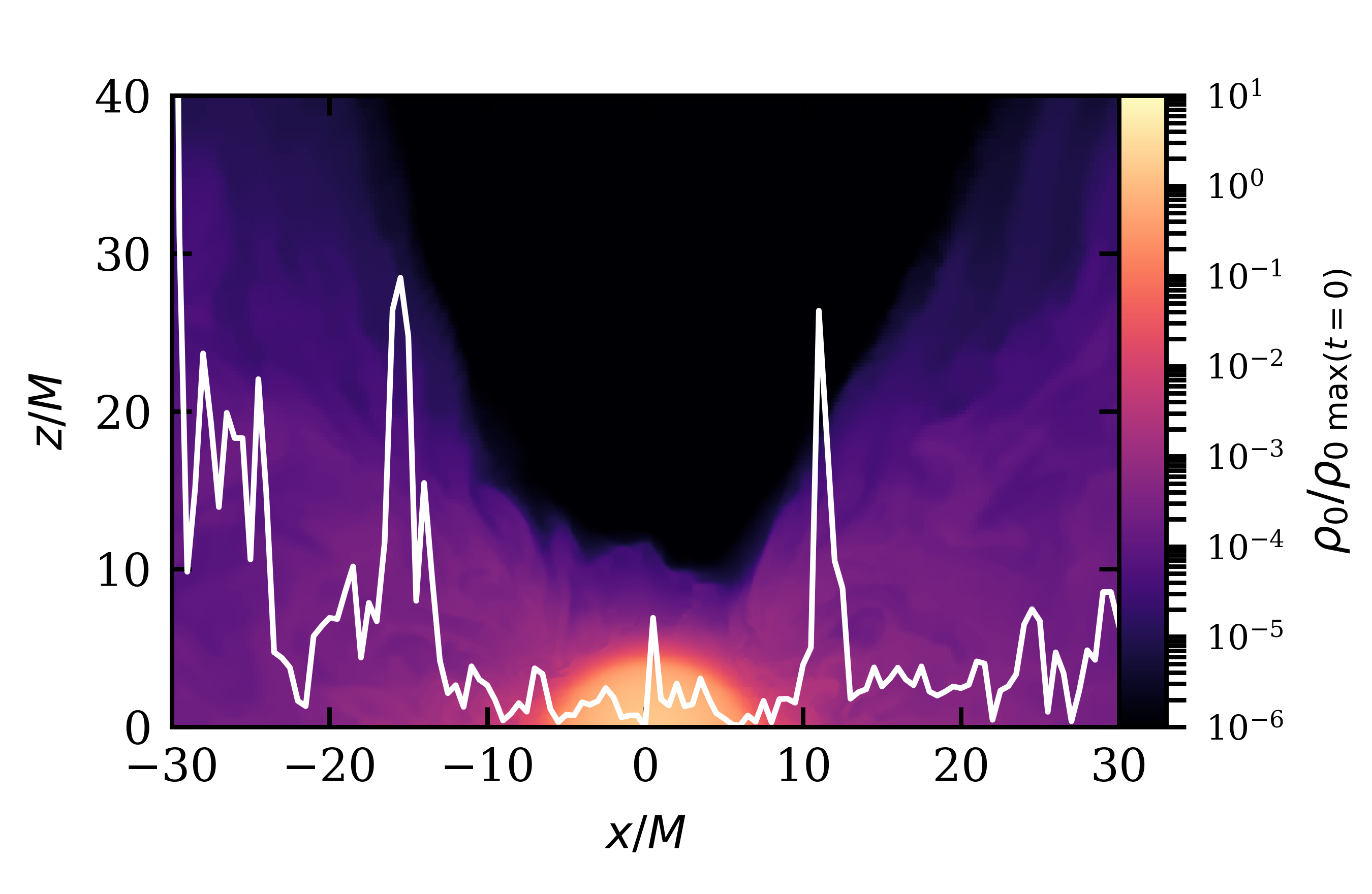}
\end{tabular}
    \caption{Top panel: Pseudocolor plot of the quality factor $Q \equiv \lambda_{\textup{MRI}}/dx$ on the equatorial plane. Bottom panel: Rest-mass density normalised to its initial maximum value on a log scale along with $\lambda_{\textup{MRI}}$ (white line) on the meridional plane. The height of the line above the $x$ axis indicates the value of $\lambda_{\textup{MRI}}$ at $x=0$. Note that almost all of the area in the top image is red, indicating that the quality factor is $Q \gtrsim 10$ as required. We see that the height of the remnant and the height of the surrounding torus is larger than $\lambda_{\textup{MRI}}$ wavelength in all but a few regions, suggesting the MRI is active. The data shown here is for the SP2.57 case depicted at $\sim 17.8$ms post merger. However, it is representative for all cases, including the HMNS remnant in case IR2.70 prior to its collapse and the post-collapse accretion disk}
    \label{fig:MRI_Q}
\end{figure}

Gottlieb et al. \cite{Gottlieb:2023est} simulate a BH+NS merger and track the outflow from the resulting BH + disk out to large distances using the endpoint (8ms post-merger) of a numerical relativity (NR) simulation of the merger itself as initial data for a GPU-accelerated MHD simulation of the resulting jet with a fixed background spacetime, reaching a total simulation time of several seconds and a distance of $\gtrsim 10^6$km $\sim 10^6 M$. However, as the NR simulation does not include magnetic fields they add seed magnetic fields by hand to the initial data for the large-scale GRMHD simulation, with various configurations and an initial ultrahigh magnetization of $\sigma_0 = 150$. They do not report an asymptotic Lorentz factor for the outflow in \cite{Gottlieb:2023est}, however in their previous large-scale GRMHD simulation of a jet produced from a BH with a magnetized torus (modelling the aftermath of a NSNS merger with delayed collapse to a BH) \cite{Gottlieb:2022sis} they show $-u_t(h + \sigma) \approx \Gamma(\sigma + h) \approx \mu$, initially $\sim 20$, is conserved out to $10^{5}$km at $t=1.7$s with efficient conversion of magnetic to kinetic energy, resulting in a measured Lorentz factor of $\sim 10$ at a distance of $10^6$km. Likewise in their simulation of the magnetorotational collapse of a massive star (a collapsar, also resulting in a BH + accretion disk) \cite{Gottlieb:2022tkb}, they show $-u_t(h + \sigma) \approx \mu$ remains $>100$ for $\sigma_0 = 200$ out to $\sim 10^3$km at 10s post-merger, beyond which it drops to $O(1)$ which they attribute to mixing between the jet material and the denser surrounding stellar cocoon. The largest Lorentz factor they observe in the collapsar simulation is $\Gamma \sim 30$. These results provide numerical evidence for efficient acceleration in outflows from BH-disk central engines - at least for the somewhat idealized initial conditions used. 
For such efficient acceleration a magnetization at the base of the outflow of $\sigma \sim b^2/\rho_0$ of $O(100)$ should lead to a asymptotic Lorentz factor of at least $\Gamma_{\infty} \sim O(10)$ up to $\Gamma_{\infty} \approx \sigma_0 \sim O(100)$.

\subsection{Magnetic amplification and instabilities}
\label{subsec:M_instabilities}

In a highly conducting plasma where magnetic field lines are ``frozen-in", winding, stretching and folding of the field lines due to shear and compression, e.g. through differential rotation or turbulent motion, can significantly amplify the field as the rotational kinetic energy of the remnant is converted to magnetic energy~\cite{Duez:2006qe,Brandenburg:2004jv,Shapiro:2000zh}. In NSNS mergers several mechanisms provide such an amplification and have been extensively studied:

\subsubsection{Kelvin-Helmholtz and Rayleigh-Taylor instabilities}

Immediately post-merger the strongest mechanism for magnetic amplification \cite{Kiuchi:2015sga,Kiuchi:2017zzg} is the Kelvin-Helmholtz instability (KHI) \cite{Chandrasekhar:1961}. For a non-synchronized binary \cite{Rasio:1999ku} when the stars come into contact they form a surface with a shear layer with fluid flowing in opposite directions on each side due to a combination of the orbital binary motion and the rotation of each star \cite{Rasio:1999ku,Obergaulinger:2010gf}. This shear interface is unstable to the KHI resulting in small perturbations growing exponentially into characteristic vortices. The resulting turbulence then amplifies the field (both poloidal and toridoidal components) via the small-scale turbulent dynamo mechanism \cite{Aguilera-Miret:2020dhz}, while also generating internal energy via viscous dissipation (numerical viscosity in our simulations). A simple Newtonian linear stability analysis suggests small perturbations should grow as $\propto \exp(\sigma_{\textup{KH}}(t - t_{\textup{merge}}))$ \cite{Chandrasekhar:1961,Schekochihin:2005rd} for two inviscid constant density fluid layers where \begin{equation}
    \sigma_{\textup{KH}} \sim \frac{\pi \Delta v}{\lambda} \sim 1000\;\textup{ms}^{-1}\left(\frac{\Delta v}{0.1 c}\right)\left(\frac{\lambda}{100 \textup{m}}\right)^{-1},
\end{equation} and where $\Delta v$ is the velocity difference across the boundary, implying that the shortest wavelength unstable modes grow fastest. For a shear layer of finite width $d$, unstable modes have wavelength $\lambda \gtrsim d$. However, in a numerical simulation the minimum wavelength and shear layer thickness are both limited by the resolution with $\lambda,d \gtrsim \Delta x_{\textup{min}}$ where $\Delta x_{\textup{min}}$ is the resolution, leading Price and Rosswog \cite{Price:2006fi} to suggest and Kiuchi et al. \cite{Kiuchi:2023obe} to find numerically that the KHI growth rate is inversely proportional to $\Delta x_{\textup{min}}$ down to $\Delta  x_{\textup{min}} \lesssim 12.5$m. 
The total magnetic energy also grows exponentially, $\propto \exp(2\gamma_{\textup{KH}}(t-t_{\textup{merge}}))$, with some characteristic growth rate $\gamma_{\textup{KH}}$ which can in principle be very different from $\sigma_{\textup{KH}}$  \cite{Rincon:2019coh}. As the unstable modes grow they enter a non-linear regime with polynomial growth rates, then eventually saturate when the magnetic field becomes large enough to oppose further distortion via Lorentz forces \cite{Schekochihin:2005rd,Rincon:2019coh} with near equipartition between magnetic and turbulent kinetic energy (magnetic energy / kinetic energy $\sim O(10^{-1})$) \cite{Skoutnev:2021chg}. Previous studies \cite{Kiuchi:2015sga,Kiuchi:2023obe} have also found that growth is terminated when the shear layer is destroyed within a few ms due to shocks and numerical viscosity. The Rayleigh-Taylor instability (RTI), which occurs when the density gradient between two fluid layers is misaligned with the local gravitational field, has also been proposed as a source of turbulence and therefore magnetic field amplification in the outer regions of the remnant \cite{Skoutnev:2021chg}, and which may complement the KHI-induced amplification which is strongest in the core \cite{Palenzuela:2021gdo}. 

\subsubsection{Magnetorotational instability}

At around $\sim 5$ms after merger, when the KHI amplification terminates~\cite{Kiuchi:2014hja}, other mechanisms, such as the MRI and magnetic winding and braking, gradually take over. The MRI \cite{Balbus:1991ay,Balbus:1998ja} occurs in any magnetized rotating astrophysical fluid whenever the angular velocity $\Omega$ decreases with radius $\partial_{\varpi}\Omega < 0$, where $\varpi$ is the cylindrical radius. Again, initial exponential growth transitions to a non-linear regime and then saturates, generating turbulence and boosting the magnetic field via a dynamo mechanism \cite{Duez:2006qe,Shibata:2006hr,Siegel:2013nrw,Kiuchi:2023obe} while transporting angular momentum from the inner to outer layers of the binary remnant which induces the formation of a central core surrounded by a Keplerian disk. 
The MRI growth rate as well as its fastest-growing wavelength are \cite{Balbus:1998ja,Shibata:2006hr,Etienne:2012te}
\begin{align}
    \sigma_{\textup{MRI}} =& \;\tfrac{1}{2} \pdv{\Omega}{\ln \varpi} \ , \label{eq:MRIsigma} \\
    \lambda_{\textup{MRI}} \approx& \frac{2\pi v_A}{\Omega} \approx  \frac{2\pi\sqrt{b^P b_P/(b^2 + \rho_0 h)}}{\Omega} \ ,
    \label{eq:MRIlambda}
\end{align}
where $v_A$ is the Alfv\'{e}n speed, $|b^P|=\sqrt{b^\mu b_\mu - (b_\mu (e_{\hat{\phi}})^\mu)^2}$ and
$(e_{\hat{\phi}})^\mu$ is the toroidal orthonormal vector comoving with the fluid.
For a Keplerian distribution, $\Omega \propto \varpi^{-3/2}$ and Eqs. (\ref{eq:MRIsigma}), (\ref{eq:MRIlambda}) 
give
\begin{align}
    \sigma_{\textup{MRI}} \sim& \;\tfrac{3}{4}\Omega = 
         1.0 \;\textup{ms}^{-1} \; \left(\tfrac{10^{3}\textup{rad s}^{-1}}{\Omega}\right), \\
    \lambda_{\textup{MRI}} \sim& \;2\textup{km}\; 
                   \left(\tfrac{10^{3}\textup{rad s}^{-1}}{\Omega}\right)
                   \left(\tfrac{B^P}{10^{15}\textup{G}}\right)
                   \left(\tfrac{\rho_0}{10^{15}\textup{g\,cm}^{-3}}\right)^{-1/2}.
\end{align}
To monitor whether we can resolve the MRI we calculate the MRI-quality factor $Q_{\textup{MRI}} := \lambda_{\textup{MRI}}/\Delta x$, where $\Delta x$ is the local grid spacing, which measures the number of grid points per wavelength of the faster growing MRI mode \cite{Ruiz:2021qmm}. Previous works \cite{Sano:2003bf,Shiokawa:2011ih} suggest we need $Q_{\textup{MRI}} \gtrsim 10$ size of remnant to properly capture the instability, and the local height of the remnant must be $>\lambda_{\textup{MRI}}$ for the instability to be active. In our simulations $Q \gtrsim 10$ across the the vast majority of the remnant and the surrounding torus (see the upper plot in Fig. \ref{fig:MRI_Q}) and the remnant and torus height is locally $>\lambda_{\textup{MRI}}$ in all but a few regions (bottom plot in Fig. \ref{fig:MRI_Q}), suggesting the MRI can operate in our remnant stars and can be captured by our simulations. Note however that the linear analysis used here assumes smooth and static background mean densities and magnetic fields, a condition that may not be satisfied in the physical remnant due to small-scale dominant variations \cite{Palenzuela:2021gdo}. 

\subsubsection{Magnetic winding and braking} 

This is not an instability but rather a secular consequence of the differential rotation and magnetic induction equation (Eq. (14) and (15) in \cite{Etienne:2010ui}) \cite{Baumgarte:1999cq,Cook:2003ku,Duez:2004nf,Spruit:1999cc,Shapiro:2000zh}. Assuming axisymmetry, a magnetic field small enough that it has negligible backreaction on the fluid, and quasiequilibrium conditions such that the fluid velocities are solely axial and slowly varying with time, we obtain (see Eqs. (2)-(7) in \cite{Duez:2006qe})
\begin{equation}
    \partial_t(\tilde{B}^{\phi}) \approx \tilde{B}^{j}\partial_i\Omega,
\end{equation}
for $j \in (\varpi,z)$ and where $\tilde{B}^{i} = \sqrt{\gamma}B^i$ and $\gamma$ is the determinant of the spatial metric. At early times the poloidal field is dominant and the toroidal field $B^T \equiv \varpi B^{\phi}$ negligible so the toroidal field grows linearly with time as \cite{Shibata:2006hr}
\begin{align}
    B^T \approx& \;t\varpi\tilde{B}^{j}(t=0)\partial_i\Omega(t=0) \sim \tfrac{3}{2}t\Omega \vert B^{\varpi}\vert, \\
    \sim& 10^{15}\textup{G} \left(\tfrac{t}{100\; \textup{ms}}\right)\left(\tfrac{\Omega}{10^3 \textup{rad s}^{-1}}\right)\left(\tfrac{\vert B^{\varpi}\vert}{10^{13}\textup{G}}\right),
\end{align}
assuming a Keplerian angular velocity profile. As the magnetic fields lines are wound up and the toroidal field increases this creates magnetic tension that acts to resist the differential rotation via magnetic braking \cite{Shapiro:2000zh}, changing the velocity profile towards $\Omega = const.$ inside the star on the Alfv\'{e}n timescale 
\begin{equation}
    t_{A} \sim \frac{R}{v_A} \sim 10\;\textup{ms} \left(\tfrac{\vert B^{\varpi}\vert}{10^{13}\textup{G}}\right)^{-1}\left(\tfrac{R}{10\;\textup{km}}\right)\left(\tfrac{\rho_0}{10^{15}\textup{g\, \rm{cm}}^{-3}}\right)^{1/2}.
\end{equation}

\subsubsection{$\alpha\Omega$ dynamo}

The growth of the large-scale magnetic field can be described through mean field dynamo theory, which relates the evolution of the mean magnetic field to the statistics of the turbulent velocity field \cite{Steenbeck:1966,Moffatt:1978,Krause:1980,Brandenburg:2004jv,Kulsrud:2005}. The key idea is that there is a separation of scales between the small scale turbulence and the large-scale magnetic field. The physical quantities $\boldsymbol{X}$ are then decomposed into average mean field $\bar{\boldsymbol{X}}$ (e.g. an azimuthal spatial average) and a fluctuating small-scale part $\boldsymbol{X}'$. The mean field induction equation is then $\partial_t \bar{\boldsymbol{B}} = \boldsymbol{\nabla} \times \left(\bar{\boldsymbol{u}} \times \bar{\boldsymbol{B}} + \bar{\boldsymbol{\mathcal{E}}}\right)$ where $\boldsymbol{u}$ is the fluid velocity and $\bar{\boldsymbol{\mathcal{E}}} = \overline{\boldsymbol{u}' \times \boldsymbol{B}'}$ is the mean electromotive force due to the fluctuations \cite{Brandenburg:2004jv,Kiuchi:2023obe}. This can be expressed in terms of the mean fields as $\bar{\mathcal{E}}_i = \alpha_{ij}\bar{B}^j + \beta_{ij}\left(\boldsymbol{\nabla}\times \bar{\boldsymbol{B}}\right)^j$. The $\alpha_{ij}$ term contributes the ``$\alpha$ effect". In particular, if we neglect the $\beta_{ij}$ term then 
\begin{equation}
    \partial_t \bar{B}^{\varpi} = -\partial_z \mathcal{E}_{\phi} \approx - \partial_z(\alpha_{\phi \phi} \bar{B}^{\phi} + \alpha_{\phi \varpi} \bar{B}^{\varpi}).
\end{equation}
The magnetic winding of the mean field is referred to as the ``$\Omega$ effect" \cite{Reboul-Salze:2021rmf} with 
\begin{equation}
    \partial_t \bar{B}^{\phi} \approx \pdv{\Omega}{\ln \varpi} \bar{B}^{\varpi},
\end{equation}
where the $\mathcal{E}$ contributions are subdominant \cite{Kiuchi:2023obe}. The combination of these two equations with a non-zero $\alpha_{ij}$ due to turbulant motion completes the $\alpha\Omega$ dynamo and allows for the amplification of both mean toroidal and poloidal fields. Kiuchi et al. \cite{Kiuchi:2023obe} have argued that it is this dynamo mechanism, powered by the MRI-driven turbulence, that creates the strong large-scale magnetic field that in turn enables the launch of a magnetically dominated jet.  

We discuss the amplification of the magnetic field energy due to the KHI, the MRI, and magnetic winding in Sec.~\ref{sec:ME_amp}. The RTI and the $\alpha\Omega$ dynamo may also be present.

\begin{table*}
\begin{ruledtabular}
\begin{tabular}{c|cccccccccccccc}
Case & $t_{\textup{GW}}$ & $t_{\textup{sim}}$ & $M_{\textup{rem}}$ & $M_{\textup{disk}}/M_0$ & $M_{\textup{esc}}$ & $\Delta E_{\textup{GW}}/M$ & $\Delta J_{\textup{GW}}/J$ & $L_{\textup{EM}}$ & $L_{\textup{fluid}}$ & $\Gamma_{\textup{max}}$ & $L_{\textup{knova}}$ & $\tau_{\textup{peak}}$ & $T_{\textup{peak}}$ & Fate. \\
\hline
IR2.40 & 8.8	& 58.58	& 2.42 & 4.36\%  &	4.33\%	& 1.81\%	& 23.8\%	& $10^{53.1}$	 & $10^{54.4}$ & 2.03	& $10^{40.94}$ & 	11.37 & 	$10^{3.14}$ &	SMNS \\
IR2.51 & 7.2	& 57.12	& 2.53 & 4.50\%  & 4.87\%	& 1.81\%	& 22.4\%	& $10^{53.0}$   & $10^{52.8}$ & 2.15	& $10^{40.91}$ & 	13.37	& $10^{3.13}$ &	SMNS \\
IR2.54 & 6.8	& 57.30	& 2.55 & 4.60\%  & 5.05\%	& 1.74\%	& 21.6\%	& $10^{53.1}$	 & $10^{52.8}$ & 1.75	& $10^{40.98}$	& 12.48 & $10^{3.12}$ &	SMNS \\
IR2.57 & 6.5	& 56.49	& 2.59 & 5.07\%  & 4.67\%	& 2.10\%	& 24.1\%	& $10^{52.9}$	 & $10^{52.7}$ & 1.71	& $10^{40.93}$	& 12.82	& $10^{3.13}$ & SMNS \\
IR2.70 & 7.0	& 57.07	& 2.52 & 2.66\%  & 2.77\%  & 2.45\%	& 25.8\%    & $10^{52.9}$	& $10^{52.7}$ & 2.13	& $10^{40.72}$	& 12.84 & $10^{3.18}$ & HMNS $\rightarrow$ BH \\
\hline
SP2.40 & 13.8	& 63.71	& 2.32	& 6.19\%	& 6.05\%	& 1.35\%	& 20.3\%	& $10^{53.2}$	& $10^{53.0}$ & 2.09	& $10^{41.01}$	& 12.96	& $10^{3.11}$	& SMNS \\
SP2.51 & 11.5	& 61.51	& 2.45		& 6.57\%	& 5.24\%	& 1.73\%	& 22.3\%	& $10^{53.2}$ & $10^{54.4}$  & 2.71	& $10^{41.02}$	& 11.81	& $10^{3.11}$	& SMNS \\
SP2.54 & 11.0 & 61.58	& 2.47	& 6.38\%	& 5.75\%	& 1.72\%	& 22.0\%	& $10^{53.2}$	& $10^{54.5}$ & 2.05	& $10^{41.03}$	& 12.80	& $10^{3.11}$ &	SMNS \\
SP2.57 &  10.5	& 60.97	& 2.53	& 5.86\%	& 5.58\%	& 1.95\%	& 23.5\%	& $10^{53.3}$	& $10^{53.1}$ & 2.06 & $10^{41.00}$ & 13.18	& $10^{3.11}$ &	SMNS 
\end{tabular}
\end{ruledtabular}
\caption{Summary of the key values for our NSNS merger simulations. Here $t_{\textup{GW}}$ and $t_{\textup{sim}}$ are the merger time at peak GW amplitude and total simulation time in ms, and $M_{\textup{rem}}$ is the rest mass of the remnants evaluated at $t_{\textup{sim}}$. The remnant mass and spin for the BH remnant (case IR2.70) is calculated using the isolated horizon formalism~\cite{Dreyer:2002mx}, while the rest mass for the NS remnants are evaluated inside a contour of $\rho_0 = 10^{-3}\rho_{0\textup{max}(t=0)}$ (our definition of stellar surface). The rest mass of the HMNS remnant for IR2.70 at $t=7.2$ms after merger (and just before BH formation) was $\sim 3.00 M_{\odot}$. The rest mass of the accretion disk / bound torus is denoted $M_{\textup{disk}}$, also evaluated at $t_{\textup{sim}}$, while $M_0$ is the initial total rest mass of the binary. $M_{\textup{esc}}$ denotes the escaping rest mass (ejecta) calculated via Eq. \eqref{eq:esc_mass} for $t > t_{\textup{GW}}$. The fractions of energy and angular momentum carried off by GWs are $\Delta E_{\textup{GW}}/M$ and $\Delta J_{\textup{GW}}/J$ respectively. $L_{\textup{EM}}$ [erg/s] and $L_{\textup{fluid}}$ [erg/s] denote the Poynting and fluid luminosity, respectively, averaged over the last $\sim$5ms of the simulation and $\Gamma_{\textup{max}}$ denotes the maximum Lorentz factor observed for the fluid \textit{within the simulation box} at $t_{\textup{sim}}$. $L_{\textup{knova}}$[erg/s], $\tau_{\textup{peak}}$[days], 
and $T_{\textup{peak}}$[K] denote the estimated peak EM luminosity in $\textup{erg}\;\textup{s}^{-1}$, rise time in days, and temperature respectively of the potential kilonova arising from the sub-relativistic ejecta. These are calculated from the ejecta mass and the ejecta velocity $v_{\textup{eje}}$ averaged over the last $\sim 500$km of the outflow. The final column shows the fate of the merger.}
\label{tab:Results_data}
\end{table*}

\begin{figure}
    \centering
    \includegraphics[width=0.45\textwidth]{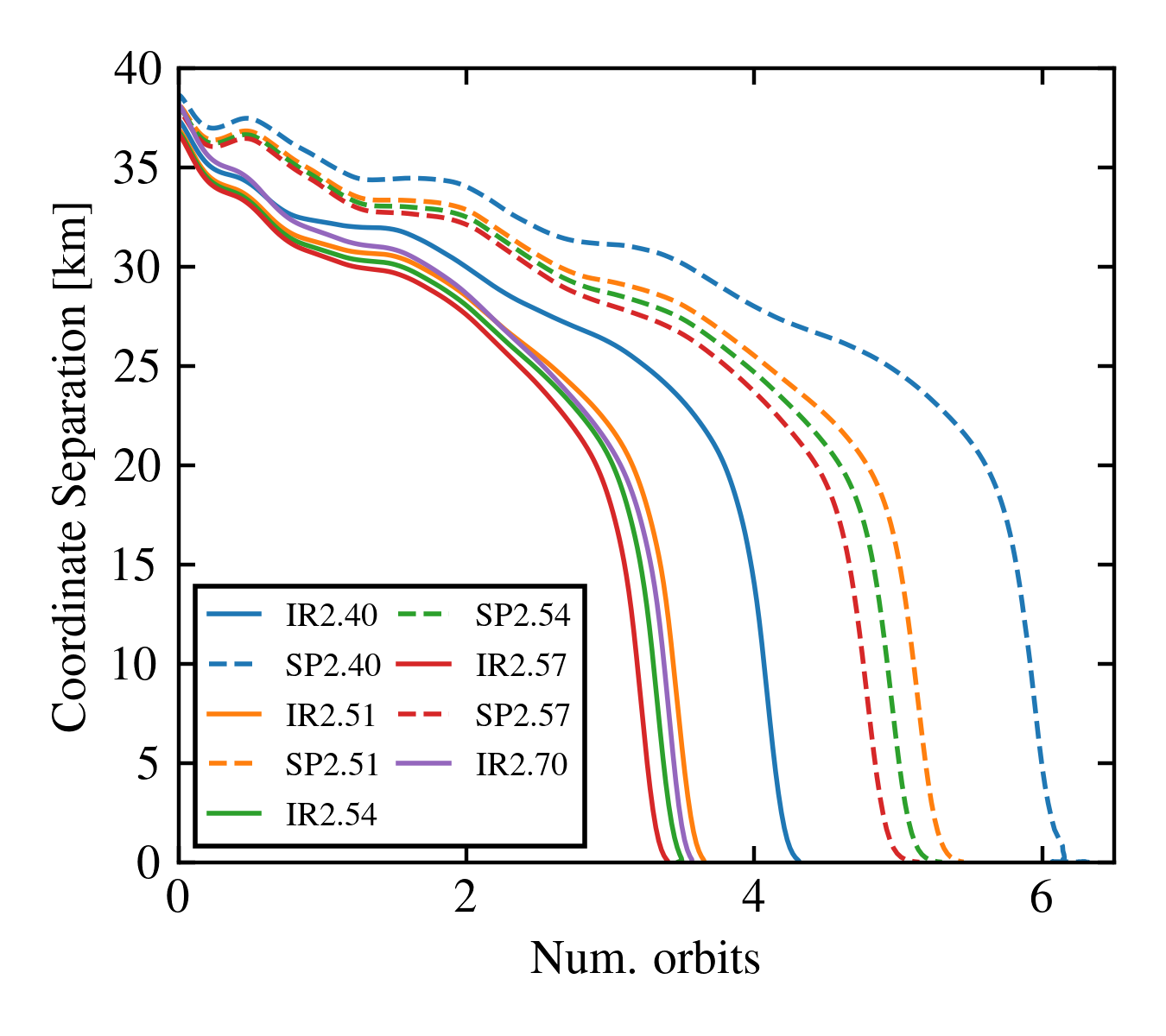}
    \caption{Binary coordinate separation of the NS centroids, defined as the position of maximum rest-mass density.}
    \label{fig:NS_sep}
\end{figure}


\section{Results}
\label{Sec:Results}

The overall dynamics of the simulations match the picture from previous works, and can be seen in 3D volume renderings of the representative IR2.54 SMNS remnant case shown in Fig. \ref{fig:3D_plots}. The stars start in quasi-circular orbits with pulsar-like poloidal dipole magnetic fields (Fig. \ref{fig:3D_plots} upper left). They inspiral due to the loss of angular momentum and energy via gravitational radiation, causing the orbital separation to shrink (see the plot of binary coordinate separation in Fig. \ref{fig:NS_sep}) until they plunge and merge. After they make contact Fig.\ref{fig:3D_plots} upper right) they form a nonaxisymmetric double core structure which oscillates as the cores collide. The outer layers gain angular momentum due to torques from the rotation of the double core structure, orbital angular momentum convection and magnetic effects. This generates a sudden outflow of ejecta as the two stars coalesce, forming two spiral tails (just visible in the lower left panel of Fig.~\ref{fig:3D_plots}) followed by a low density torus of matter surrounding the central remnant (\ref{fig:3D_plots} lower right). The supramassive remnant continues to lose energy and angular momentum via gravitational waves until it becomes axisymmetric and quasistationary approximately $30$ ms after merger. 

\begin{figure*}
    \centering
    \includegraphics{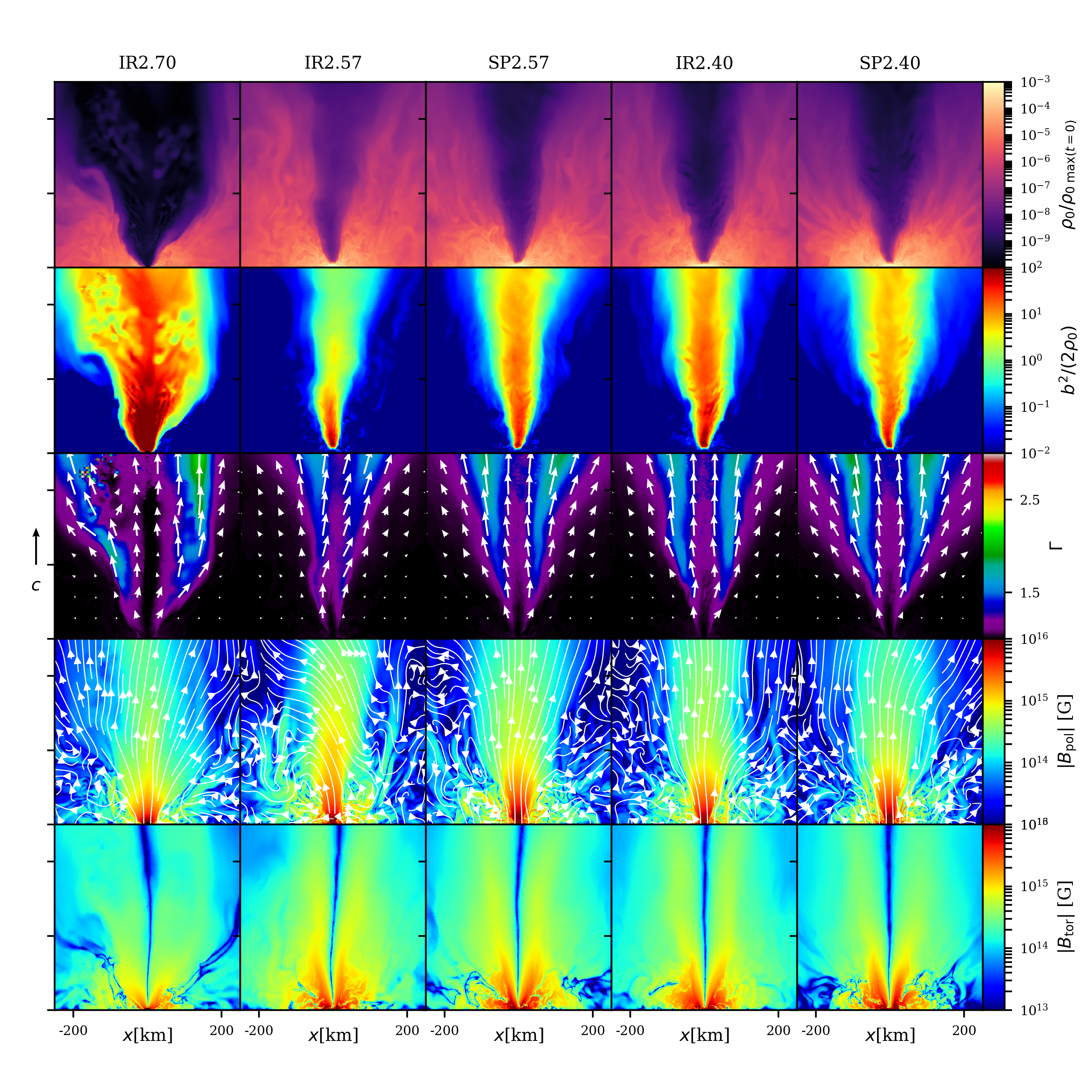}
    \caption{2D plots in the meridional plane at 20ms post-merger for five of the simulations in Table~\ref{tab:initial_NS}. We show the rest-mass density $\rho_0$ on a log scale relative to the initial maximum rest-mass density (top row), $b^2/(2\rho_0)$ which approximately corresponds to the magnetization $\sigma$ (second row), the Lorentz factor (third row), and the strength of the poloidal and toroidal magnetic field (fourth and fifth rows), with the magnetic field lines shown in the poloidal field plots. The scale on the two axes is the same. The arrows in the Lorentz factor plot (third row) indicate the flow velocities, and the arrow labelled on the left-hand side indicates the magnitude of the speed of light.}
    \label{fig:2D_plots}
\end{figure*}

The magnetic field is amplified and a strong toroidal field is generated, mainly due to magnetic winding from the differential rotation. The magnetic pressure and Poynting flux is able to overcome the ram pressure to accelerate and clear away the gas in the polar regions, forming a evacuated low density magnetically dominated ($b^2/(2\rho_0) \gg 1$) funnel. This in turn enables the launch of an incipient jet-like structure: a mildly relativistic outflow along that funnel, confined with a tightly wound helical magnetic field from the poles (the outflow and helical field lines are also shown in Fig.~\ref{fig:3D_plots} lower right). A summary of the physical properties of the NSNS remnants and their final fate is shown in Table ~\ref{tab:Results_data}.

\begin{figure*}[th]
    \centering
    \includegraphics[width=\textwidth]{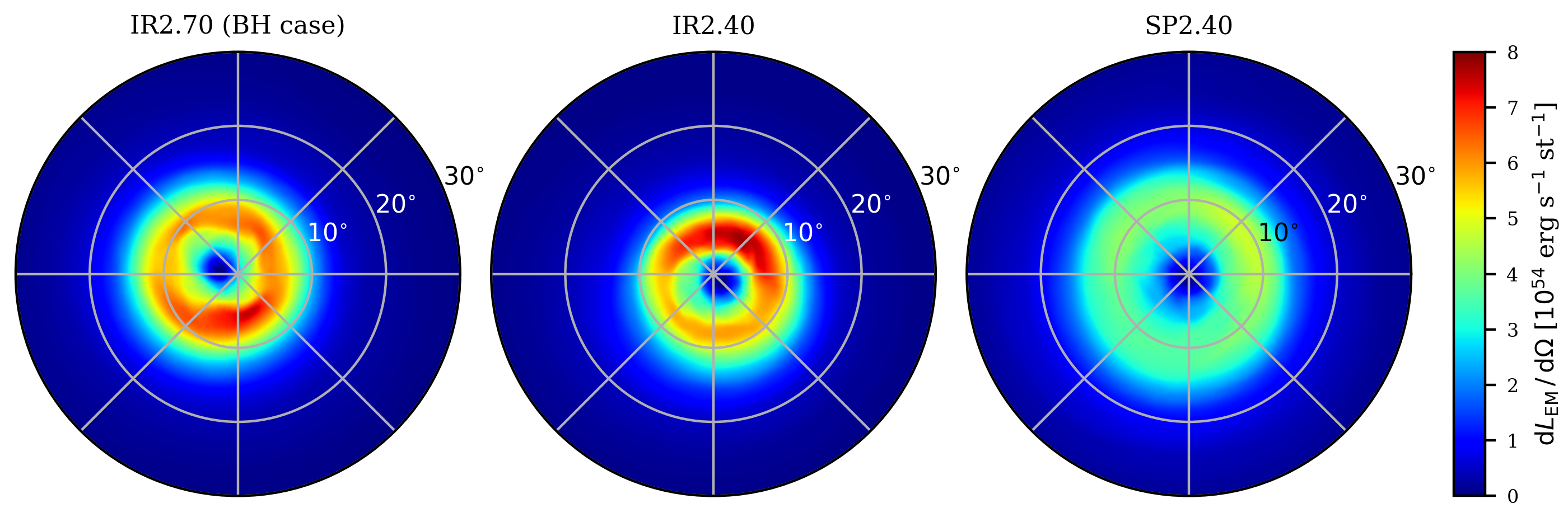}
    \caption{Angular distribution of electromagnetic Poynting luminosity per solid angle for three cases at a distance $r = 1024$km and time $t - t_{\textup{merge}} \sim 17$ms for $\theta \leq 30^{\circ}$, where $\theta$ is the polar angle from the $z$ axis (here ``$\textup{st}$" denotes ``steradian").}
    \label{fig:EM_angle}
\end{figure*}

\subsection{IR2.70: HMNS/BH remnant}
\label{sec:HMNS}

The behaviour of the IR2.70 case closely matches the very similar SLyM2.7P simulation reported in Ruiz et al. (2021) \cite{Ruiz:2021qmm}. The stars collide to form a differentially rotating remnant with an initial total rest mass of $\sim 3.0M_{\odot}$. As this is above the supramassive limit (see Table~\ref{tab:SLy_mass_lims}) the remnant is a transient HMNS which oscillates and, after a lifetime of $\sim 7.2$ms, collapses to form a spinning BH, which at the end of the simulation has a gravitational mass of $2.52M_{\odot}$ and a dimensionless spin of $\chi = 0.64$. The HMNS lifetime is slightly shorter than the $\sim 12$ms reported in \cite{Ruiz:2021qmm}; however, the initial magnetic field strength used in that study was $10^{15.5}$G, a factor of $\sim 2$ larger than in this present study. A lower magnetic field leads to a lower pressure making the HMNS remnant less stable against collapse compared to the previous case. We also note that the sensitivity of the collapse time to the magnetic field is physical and well known~\cite{Ruiz:2016rai,Giacomazzo:2010bx}. Following collapse, we see the formation of a magnetized accretion disk surrounding the BH and the continued winding up of the magnetic field lines into a tightly coiled helical shape.

Accretion into the BH down the polar axis and the acceleration of gas outwards due to the magnetic forces produces a low density ($\rho_0 \;\lesssim 10^7\textup{g}\,\textup{cm}^{-3}$) evacuated funnel\footnote{Notice that in our previous NSNS merger simulations where neutrinos were incorporated using the truncated-moment (M1) formalism the baryon pollution in the funnel was a factor of $\sim 10$ lighter than in this case~\cite{Sun:2022vri}.} with a half-opening angle at the base, defined in terms of the angular width of the magnetically dominated region, of $\sim 30^{\circ}$ (see the top left image of Fig. \ref{fig:2D_plots}). The half-opening angle inferred from distant electromagnetic observations is significantly smaller, as at large distances the outflow approaches cylindrical flow \cite{Vlahakis:2003si}, with an angle $\sim 10^{\circ}$ at distance $r = 1024$km (see Fig. \ref{fig:EM_angle} left panel).

\begin{figure}
    \centering
    \includegraphics{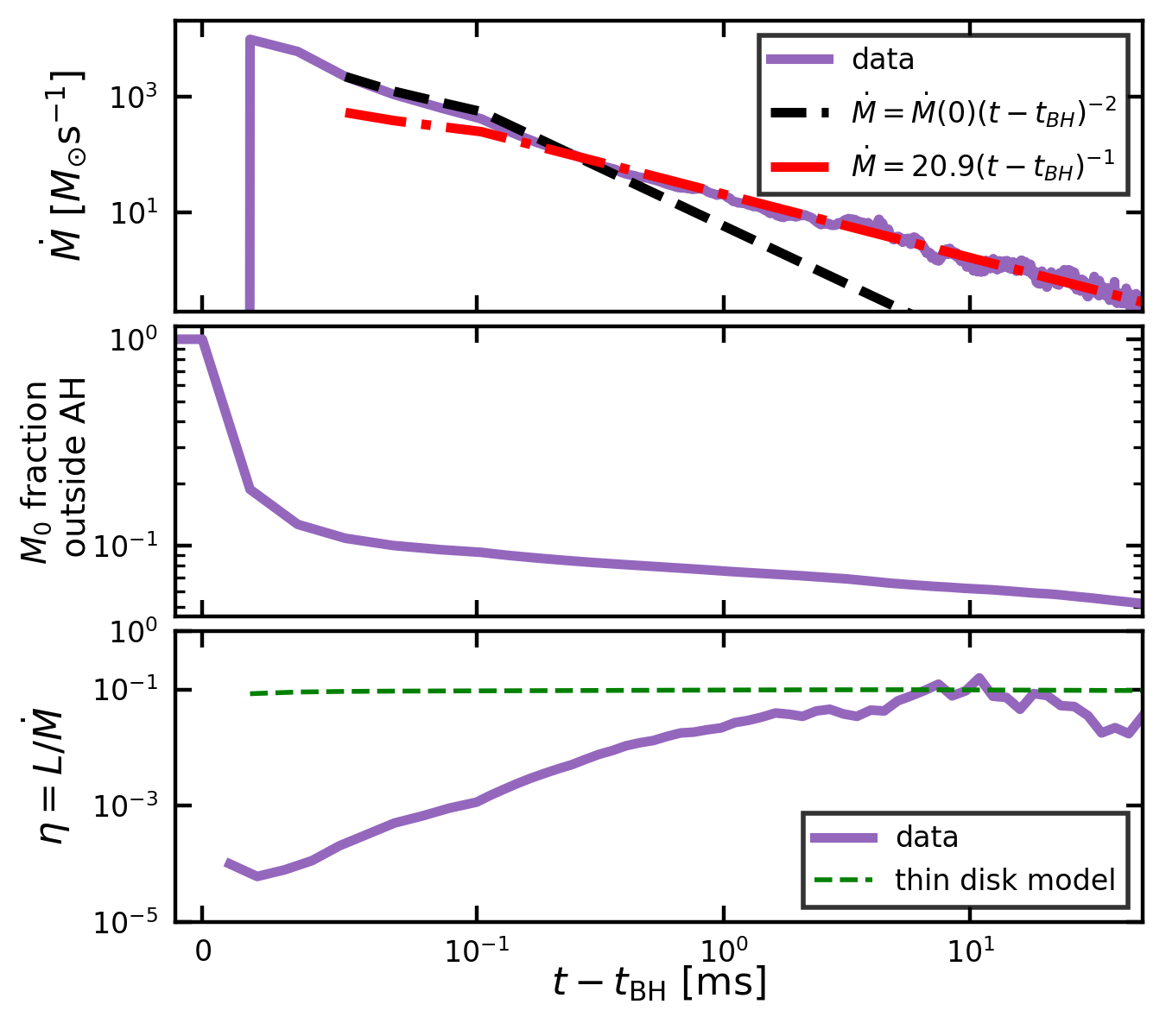}
    \caption{The accretion rate onto the BH remnant for the IR2.70 case (top), the fraction of the total rest mass outside the apparent horizon (middle), and the jet launching efficiency ($\eta = L/\dot{M}$ where $L = L_{\textup{EM}} + L_{\textup{fluid}}$ is the total jet luminosity) on log scales. The accretion rate decays as $\dot{M} \propto t^{-2}$ in the first $\sim 0.5$ms post-BH formation, as expected for highly magnetized disks \cite{Gottlieb:2023est}, then rapidly transitions to $\dot{M} \propto t^{-1}$ as for hydrodynamically-dominated accretion, where accretion is driven by shocks and spirals in the disk \cite{Gottlieb:2023est}. We also show the expected accretion efficiency for a Novikov-Thorne thin disk model (green dashed line, lower panel).}
    \label{fig:BH_accretion}
\end{figure} 

The low density and strong $\sim 10^{16}$G magnetic field enables a force-free environment with $b^2/(2\rho_0) \sim \sigma \gg 1$ in the funnel above the BH poles, and the helical field enables the acceleration of an unbound, collimated, mildly relativistic ($\Gamma \sim 2.0$ in our simulation box) outflow, mainly along the outer regions of the funnel. This behaviour is consistent with the standard picture of Poynting-flux driven outflow reported in the literature (e.g. \cite{McKinney:2008ev}) which we describe as an incipient jet. The accretion rate (see Fig. \ref{fig:BH_accretion} top panel) shows a power-law decay with time, initially falling with $\dot{M} \propto t^{-2}$ in the first $\sim 0.5$ms, consistent with the expectation for a magnetically dominated disk \cite{Gottlieb:2023est}, then rapidly transitions to $\dot{M} \propto t^{-1}$ for the remaining simulation time, consistent with a hydrodynamically-dominated disk where accretion is driven by shocks and density spirals \cite{Gottlieb:2023est}. The accretion efficiency (see Fig. \ref{fig:BH_accretion} bottom panel), $\eta := (L_{\textup{fluid}}+L_{\textup{EM}})/\dot{M}$, increases to $\sim 0.09$, close to the value expected for the thin disk Novikov-Thorne model (see \cite{Shapiro:2004ud}), which has shown remarkably close agreement with numerical MHD simulations of BZ jets \cite{McKinney:2004ka,Shapiro:2004ud}, but below the maximum value for a magnetically-arrested disk (MAD)  \cite{Lowell:2023kyu}. From the $\dot{M} \propto t^{-1}$ accretion rate we can estimate the lifetime of the accretion disk as $\sim 2.2$s, very close to the typical lifetime of a sGRB, although our lack of knowledge of the precise details of how the energy from the incipient jet is converted to the $\gamma$-ray signal we observe limits the usefulness of this comparison. 

\begin{figure}
    \centering
    \includegraphics{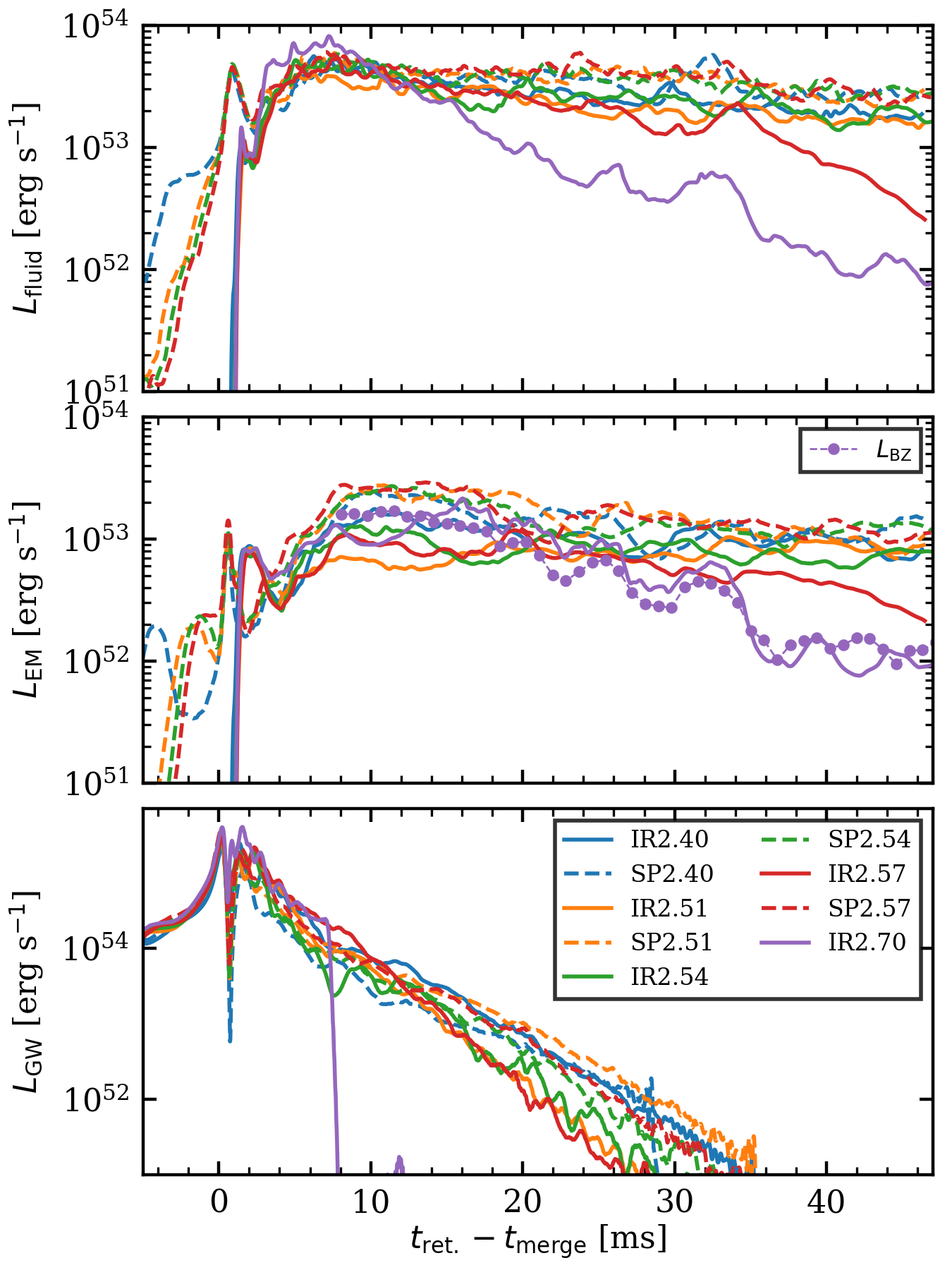}
    \caption{Luminosity via fluid outflow, EM radiation and gravitational waves, vs. retarded time since $t_{\textup{merge}}$ (defined as the time of peak GW emission) extracted over a sphere at $r_{\textup{ext}} = 1024\textup{km} \sim 260M$. The purple dots and connecting line in the middle plot show the expected BZ electromagnetic luminosity from Eq. \eqref{eq:BZ_lum} for the BH mass, spin and approximate $B_{\textup{pol}}$ for case IR2.70.}
    \label{fig:luminosity}
\end{figure}

Following the onset of collapse, the bulk of the HMNS is swallowed by the BH leaving an accretion disk which initially has $\sim 10\%$ of the initial rest mass of the system, dropping to $2.66\%$ (Table \ref{tab:Results_data} and Fig. \ref{fig:BH_accretion} middle panel) by the end of the simulation. The accretion of magnetized gas causes the total magnetic energy to drop, with a consequential decline in the electromagnetic luminosity over time. The expected electromagnetic luminosity for a jet powered by the steady-state BZ mechanism can be estimated as \cite{Thorne:1986} 
\begin{equation}
    L_{\textup{BZ}} \sim 10^{52} \left(\frac{\chi}{0.64}\right)^2\left(\frac{M_{\textup{BH}}}{2.5M_{\odot}}\right)^2\left(\frac{B_{\textup{pol}}}{10^{16}\textup{G}}\right)^2\textup{erg}\,\textup{s}^{-1},
    \label{eq:BZ_lum}
\end{equation}
where $B_{\textup{pol}}$ is the poloidal magnetic field measured at the BH pole. The luminosity from Eq. \eqref{eq:BZ_lum} is plotted as connected dots in the middle panel of Fig. \ref{fig:luminosity}. We can see that the EM luminosity for the IR2.70 case converges towards the $L_{\textup{BZ}}$ value towards the end of our simulation as the additional power from the magnetized torus dies down and the BZ mechanism dominates. The EM luminosity magnitude lies within the narrow ``universal range" of $10^{52\pm1}$erg s$^{-1}$ derived in \cite{Shapiro:2017cny} for accreting BH + disk systems resulting from compact binary mergers containing NSs, as well as from the magnetorotational collapse of massive stars. This, combined with the other criteria already validated for the very similar case reported in Ruiz et al. 2021 \cite{Ruiz:2021qmm}, demonstrates that the BZ mechanism is operating in our simulation, as we concluded in \cite{Ruiz:2016rai,Ruiz:2017due,Paschalidis:2014qra,Sun:2022vri}. One can relate the total electromagnetic luminosity $L_{\textup{EM}}$ observed in our simulations to the observed isotropic-equivalent $\gamma-$ray luminosity for a jet observed head-on as
\begin{equation}
    L_{\gamma,\textup{iso}} = \frac{1}{C_{\textup{col}}}\eta_{\textup{EM}}L_{\textup{EM}},
\end{equation}
where $C_{\textup{col}}$ is a factor that accounts for $\gamma$-ray collimation \cite{Khan:2018ejm}, and $\eta_{\textup{EM}}$ corresponds to the efficiency of converting the outgoing Poynting flux in the simulation region to $\gamma-$ray photons in the emission zone. For perfect collimation, where all the emission is contained within a homogenous jet of half-opening angle $\theta$, we have $C_{\textup{col}} = 1 - \cos \theta \sim 10^{-2}-10^{-1}$ for $\theta \sim 10^{\circ}-30^{\circ}$ \cite{Escorial:2022nvp}. 

The prompt emission mechanism is still an open question \cite{Parsotan:2022uur} making the efficiencies highly uncertain \cite{Beniamini:2016hzc}, although $\sim 10^{-1}$ has been taken as a fiducial value \cite{Khan:2018ejm} for the EM Poynting luminosity for outflow from a black hole central engine. For our cases with a SMNS central engine we see indications of mixing between the central funnel and the debris torus, associated energy loss from the outflow, and larger baryon pollution compared to the black hole central engine case, which may indicate that the efficiency in these cases is $\ll 10^{-1}$. Taking $\eta_{\textup{EM}} \sim 10^{-2} - 10^{-1}$ we get an estimated $L_{\gamma,\textup{iso}} \sim 10^{52}-10^{54}$ erg s$^{-1}$. The lower end of this estimate is compatible with sGRB observations \cite{Li:2016pes,Shapiro:2017cny,Beniamini:2020adb}. With regards to the higher end of the $L_{\gamma,\textup{iso}}$ estimate, it is possible that the large luminosity is due to the artificially strong magnetic field used, or the choice of magnetic field geometry, although as discussed in Sec. \ref{sec:Intro}, previous works have found that a strong large-scale poloidal magnetic field is a requirement for jet launching \cite{Ruiz:2020via,Etienne:2012te}. Hence, our simulations represent a ``best-case" scenario for jet formation. We intend to investigate the impact of magnetic field strength and topology in a future work.

Although we only observe a mildly relativistic outflow within our simulation region, an extensive literature suggests that the BZ effect alone can ultimately accelerate gas to ultrarelativistic ($\Gamma_{\infty} \gg 100$) velocities, even with significant baryon pollution (see e.g. \cite{Paschalidis:2014qra,McKinney:2005zx}) and therefore provide the central engine for a sGRB-compatible jet. 

\subsection{SMNS remnants}
\label{sec:SMNS}

\begin{figure}
    \centering
    \includegraphics{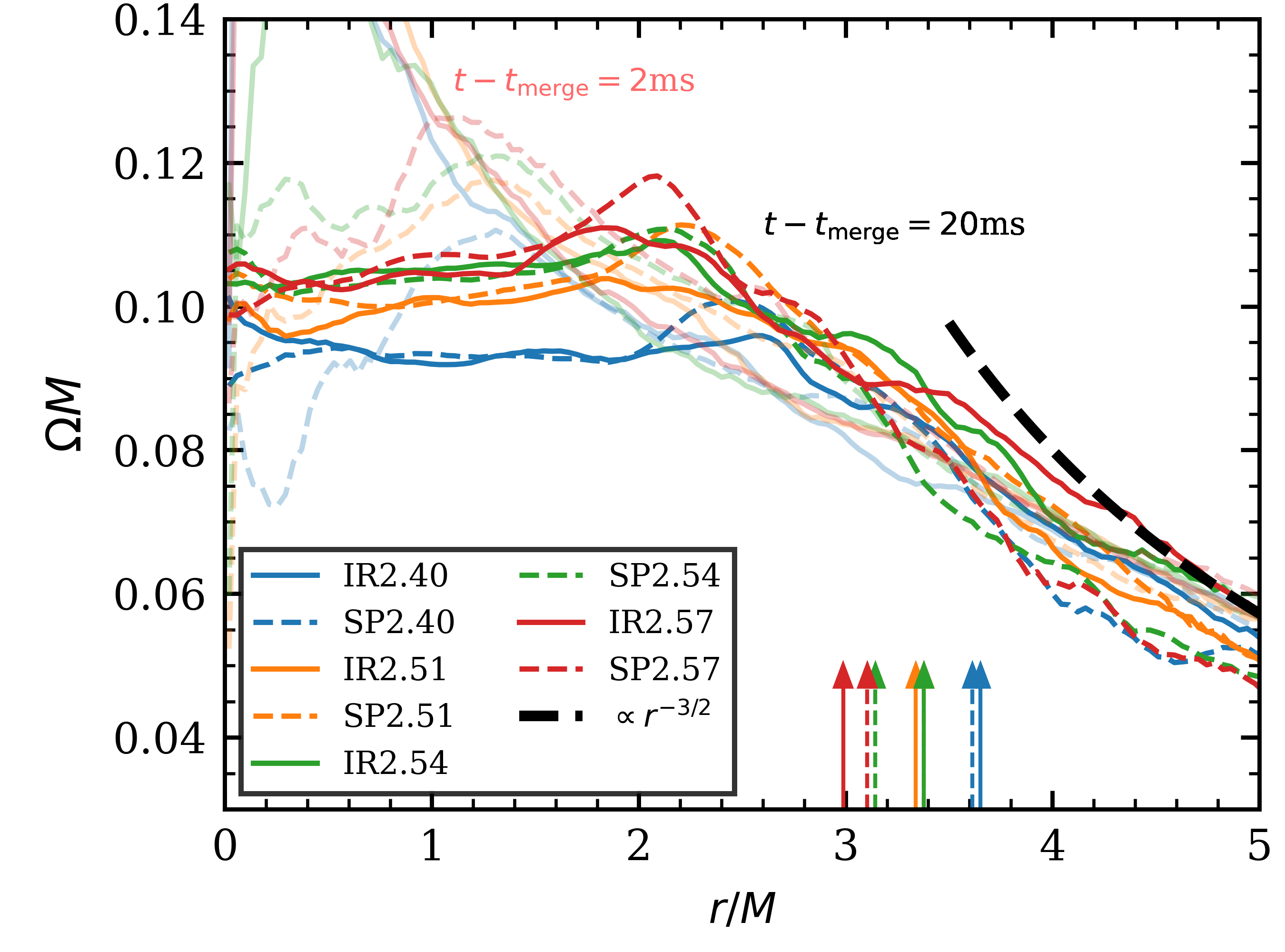}
    \caption{Rotation profiles of the SMNS remnants at 2ms post merger (faded lines) and at 20ms (bold lines). We also show a curve (thick black dashed line) for a Keplerian $\Omega \propto r^{-3/2}$ rotation profile, and arrows indicating the radii where $\rho_0/\rho^{\textup{max}}_0 = 10^{-2}$, which gives a rough measure of the size of the remnants.}
    \label{fig:Omega_curves}
\end{figure}

Having established a benchmark with the IR2.70 case with a HMNS remnant that undergoes delayed collapse to a BH, we can now consider the lower mass cases which form longer lived SMNS remnants, the main focus of this work. The first part of the evolution for these cases is very similar to the IR2.70 case, except that the outcome is a SMNS which lasts until the end of the simulation unlike the transient HMNS. The stars are driven to uniform rotation (see Fig. \ref{fig:Omega_curves}) by magnetic turbulent viscosity as angular momentum is redistributed from the inner to outer regions, forming a central uniformly rotating core and an outer low-density torus with a Keplerian rotation profile. The mass of the bound material in the disk/torus is $\sim 4-6\%$ of the initial rest mass by the end of the simulation, significantly larger than the $2.66\%$ in the BH case.   

\begin{figure}
    \centering
    \includegraphics{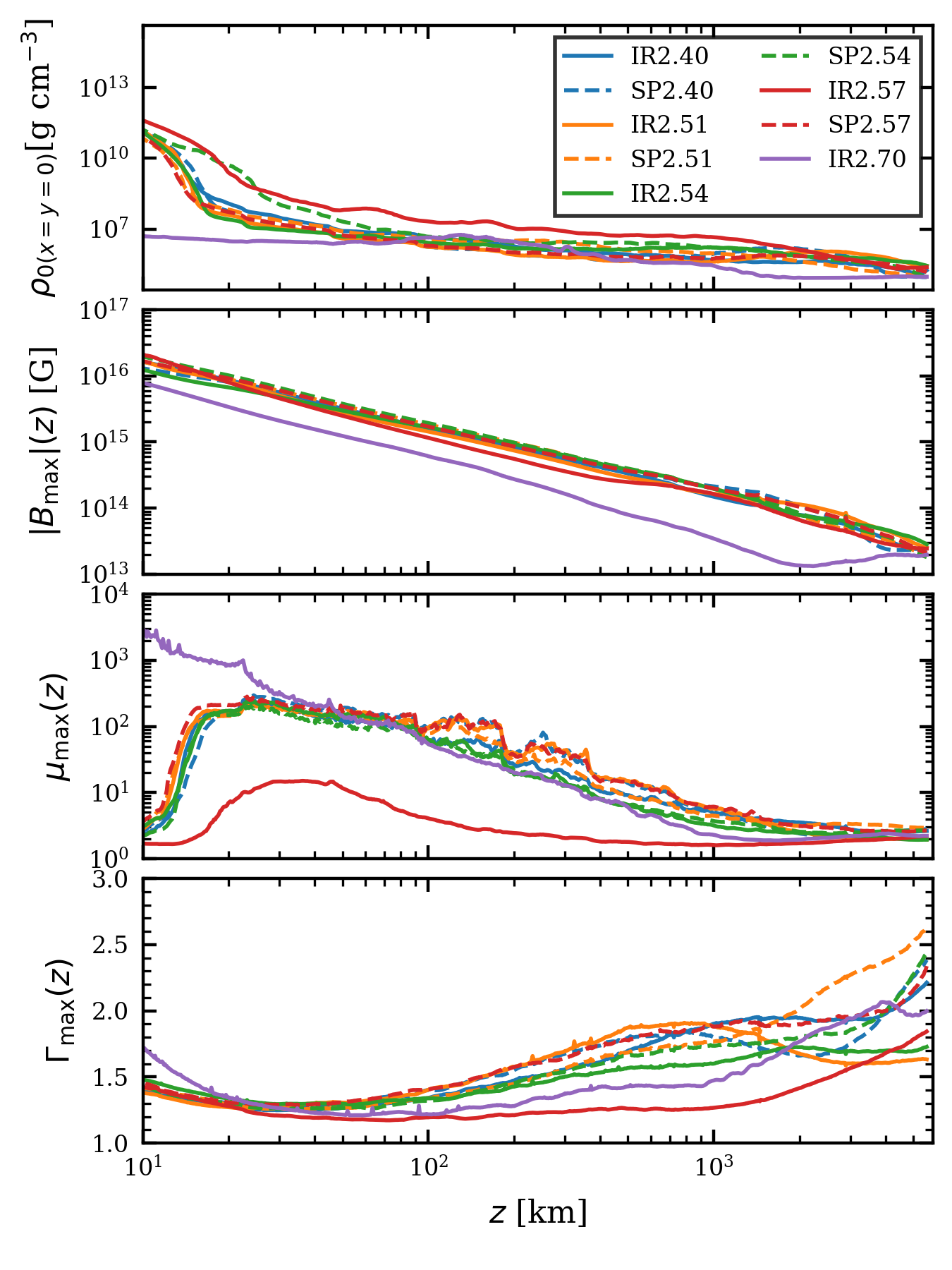}
    \caption{Rest-mass density along the $z$ axis (top), maximum comoving magnetic field $\vert B_{\textup{co}}\vert = \sqrt{4\pi b^2}$ at each height (second from top), maximum energy to mass flux ratio $\mu$ at each height (second from bottom) and maximum Lorentz factor at each height (bottom). The values are averaged over the last $10$ms of each simulation up to $50$ms post-merger.}
    \label{fig:z_axis_plot}
\end{figure}

The BZ mechanism cannot operate without an ergosphere \cite{Ruiz:2020zaz}, however the rotation of the magnetized stars twists the frozen-in magnetic field lines in much the same fashion, forming same kind of tightly wound helical magnetic structure seen in the IR2.70 case (see Fig. \ref{fig:3D_plots} lower right). The Poynting flux generated by the rotating coiled magnetic field accelerates an outflow wind of gas, again forming an evacuated funnel, although unlike the BH case, there remains a high density region of $\rho_0 \sim 10^{11}\textup{g}\,\textup{cm}^{-3}$ at the base of the funnel extending to $\sim 20$km above the NS poles (see Fig. \ref{fig:MRI_Q} bottom panel, Fig. \ref{fig:2D_plots} top row and Fig. \ref{fig:z_axis_plot} top panel). The funnel itself is also narrower (half-opening angle of $\sim 25^{\circ}$ at the base) and much more polluted with baryonic gas than for the IR2.70 BH case (Fig. \ref{fig:2D_plots} top row). 

The magnetic field is actually slightly stronger than in the BH case (see bottom two rows Fig. \ref{fig:2D_plots} and the second from top panel of Fig. \ref{fig:z_axis_plot}). This may be because there remains a much larger amount of highly magnetized material compared to the IR2.70 case where such material is rapidly accreted by the BH. However, the increased baryon pollution means that the $b^2/(2\rho_0) \sim \sigma$ is nonetheless lower than in the IR2.70 BH case (second row Fig. \ref{fig:2D_plots}), although still $\gg 1$ in the funnel center close its base. We also see a collimated, mildly relativistic outflow ($\Gamma \sim 2.0$ within in our simulation box). 

From $\sim 2$ms post-merger up $\sim 7.2$ms  when the HMNS collapses, the IR2.70 case has a slightly larger fluid luminosity than the SMNS cases. However from $\sim 7.2$ms onwards the SMNS have a larger $L_{\textup{fluid}}$ and $L_{\textup{EM}}$, which is likely because the SMNS retain a larger magnetic field and a larger reservoir of gas. For all the SMNS cases but the IR2.57 case, which is particularly choked with baryon pollution, the EM luminosity is still $\sim 10^{53}$ erg s$^{-1}$ by the end of the simulation (see Table \ref{tab:Results_data} and Fig. \ref{fig:luminosity}). 

Although we see jet-like structures in the SMNS cases that meet our basic criteria for an ``incipient jet" with mildly relativistic outflow, tightly wound helical magnetic fields, and $b^2/(2\rho_0) \gtrsim 1$ (at least for some regions of the central funnel), we \textit{cannot} conclude that these outflows can produce the true ultrarelativistic jets at large distances needed for sGRBs. 

While semi-analytic studies of idealized MHD jets suggest that $\mu$ should be conserved along a poloidal flow line, we see the maximum $\mu$ decrease with distance along the outflow funnel in all our simulations (Fig. \ref{fig:z_axis_plot} second panel from bottom), suggesting that energy is lost via mixing between the low density funnel material and the surrounding denser torus and the incipient jet is choked before it can reach ultrarelativistic speeds. While this non-conservation of $\mu$ could be an artifact resulting from decreasing numerical resolution further from the central object, it could also be a physical consequence of moving away from the idealized axisymmetric, steady-state, Minkowski spacetime model considered in semi-analytic works. The numerical results for a BH + torus central engine reported in Gottlieb et al. \cite{Gottlieb:2022tkb,Gottlieb:2022sis} show conservation of $\mu$ out to $\sim 10^5$km with a full 3D evolution. However, they use idealized axisymmetric initial conditions and as such they may be much closer to the idealized models considered in the semi-analytic literature. Gottlieb et al. also note that low baryon pollution in the outflow funnel is a requirement for jet break-out.

For BZ-powered outflows from BHs with accretion disks there is nonetheless a strong body of evidence that they can efficiently accelerate gas and reach ultrarelativistic speeds, even with baryon loading \cite{Paschalidis:2014qra,McKinney:2005zx}. Therefore we feel confident that the ``incipient jet" outflow for cases with a BH central engine, including the IR2.70 case considered here, can be the projenitor for the ultrarelativistic jet needed to produce a sGRB, despite the apparent decline in $\mu$. However, for outflow powered by a rotating magnetized star we do not have the same body of evidence, or large-scale MHD simulations such as \cite{Gottlieb:2022tkb,Gottlieb:2022sis}, and therefore cannot be confident can it can efficiently convert magnetic energy to kinetic and accelerate the gas to ultrarelativistic speeds in practice. We can only note that we observe \textit{jet-like structures}, which meet our basic criteria for an \textit{incipient} jet, and leave the question of whether these can break out and produce true ulrarelativistic jets and source sGRBs for further study (potentially with MHD simulations that extend the outflow to much larger $z \sim 10^6 $km spatial scales).  

From our preliminary study using a M1 scheme for neutrino radiation \cite{Sun:2022vri} we expect that the inclusion of neutrinos will somewhat improve the picture for SMNS jets, decreasing the baryon density in the funnel by a factor of $\sim 10$ for BH central engine incipient jets. However, it is unlikely that neutrino processes will change the overall outcome in the SMNS scenario because they have the largest impact in baryon-poor environments \cite{Just:2015dba}.

\subsection{Effects of SMNS mass}

Among the SMNS remnants the mass of the remnant does not have a very significant effect on the central outflow. The highest mass irrotational SMNS case, IR2.57, has a factor of $\sim 10$ more baryon pollution in the outflow funnel region compared to the lower mass irrotational cases, and consequently a lower $b^2/(2\rho_0)$, a smaller luminosity and a smaller outflow Lorentz factor $\Gamma$. However, the differences between the highest and lowest mass spinning SMNS cases are small (see Fig. \ref{fig:2D_plots}). 

We see that the torus of bound debris material becomes much more compact with higher remnant masses, and more diffuse with lower masses, as one would expect due to the decreased gravitational attraction of the central object. The total mass of the bound debris disk increases with mass, as a proportion of the total rest mass, for the irrotational cases despite becoming more compact (see Table \ref{tab:Results_data}). 

\subsection{Effects of initial NS spin: Comparison between spinning and irrotational cases}

The first obvious impact of initial NS spin is that it delays the inspiral of the stars by $\sim 2$ orbits (Fig. \ref{fig:NS_sep}): the well-known ``hang-up" effect due to spin-orbit coupling \cite{Campanelli:2006uy,Tsokaros:2019anx}. The additional angular momentum in the system due to the spin of the initial NSs results in larger amounts of dynamical ejecta for the same initial mass and a larger and more diffuse debris disk (see Table \ref{tab:Results_data}). The EM luminosity prior to merger is substantially larger for the spinning cases than for the irrotational cases, as the frozen-in magnetic field is subject to more winding. Following merger, the EM luminosities remain consistently larger for spinning compared to irrotational cases, although only by a factor of $\sim 1.25-2$. Our results also show that the half-opening angle of the Poynting luminosity per solid angle at large distances is consistently slightly larger for the spinning vs. the irrotational cases (see Fig. \ref{fig:EM_angle} center and right panels for the representative $M=2.40 M_{\odot}$ cases), giving a tantalizing hint that the additional angular momentum from spinning progenitors may alter the outflow morphology. We hope to explore this further in a future work. Finally, we see that the post-merger higher frequency component of the GW signal is supressed in the spinning cases compared to the irrotational cases (see Sec.~\ref{sec:GW}). 

\subsection{Ejecta mass and kilonova estimates}

\begin{figure}
    \centering\includegraphics{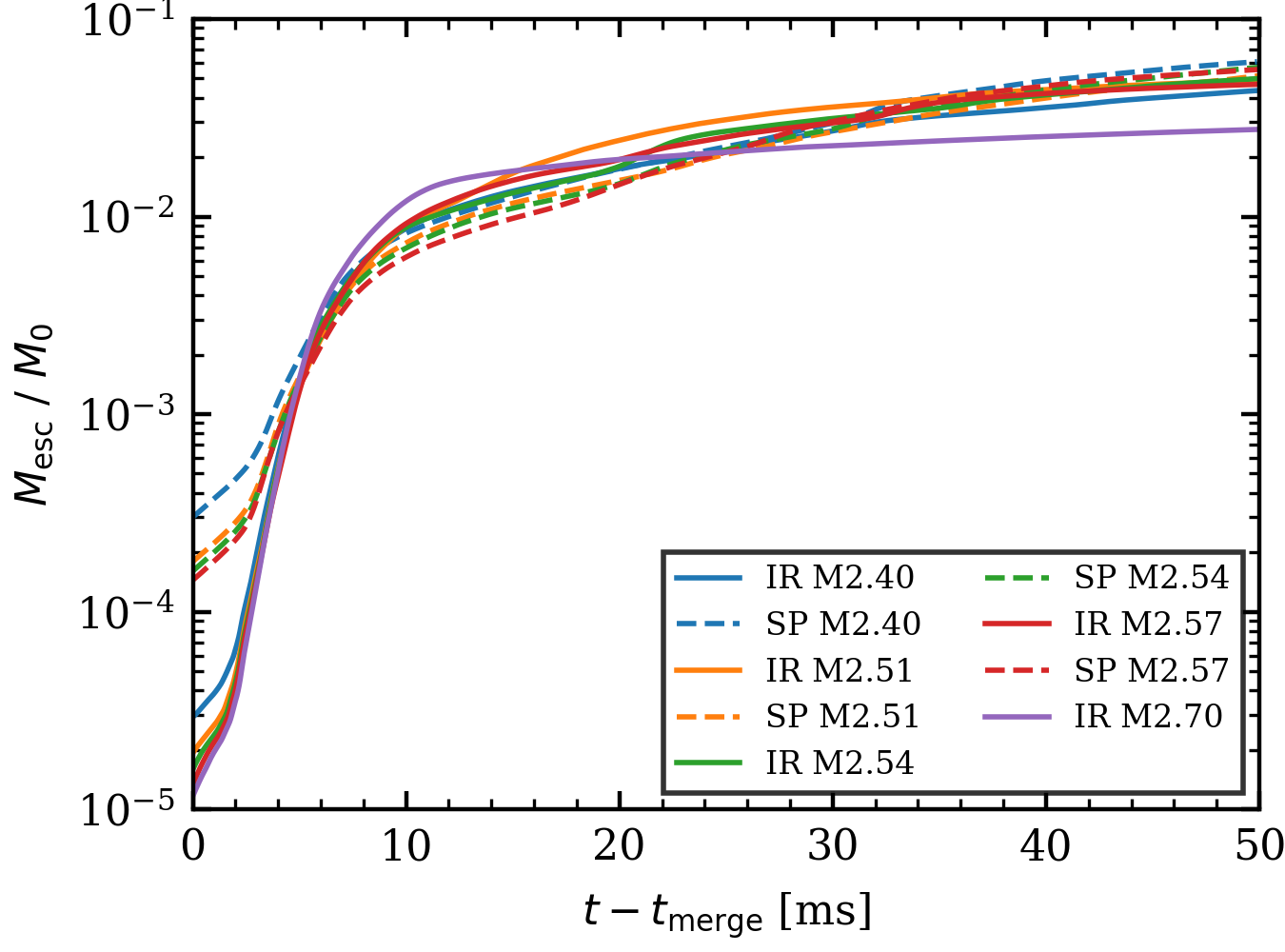}
    \label{fig:M_esc}
    \caption{Escaping rest mass $M_{\textup{esc}}$ divided by the initial total rest mass $M_0$.}
    \label{Fig:escaping}
\end{figure}

As shown in Fig.~\ref{Fig:escaping}, in all the cases we see the dynamical ejection of unbound material with an ejecta mass fraction of at least $\sim 2\%$, from $2.77\%$ for the IR2.70 HMNS case to between $4-6\%$ for the SMNS cases, corresponding to $\gtrsim 0.05 M_{\odot}$, with average velocity $\langle v_{\textup{eje}}\rangle \sim 0.2$c. This quantity of ejected matter would be expected to produce a kilonova transient (see e.g. \cite{Metzger:2019zeh}), as discussed in Sec.~\ref{sec:Intro}. Using the analytic model derived in \cite{Perego:2021dpw} and discussed in further detail in \cite{Ruiz:2021qmm}, Sec. III C (see Eq. (8)-(10) in \cite{Ruiz:2021qmm}) we estimate the peak luminosity $L_{\textup{knova}}$, the peak time of the kilonova emission $\tau_{\textup{peak}}$ and the effective black-body temperature $T_{\textup{peak}}$. We find $L_{\textup{knova}} \sim 10^{41}$erg, $\tau_{\textup{peak}} \sim 11-13$days, and $T_{\textup{peak}}\sim 1300 $K (full details shown in Table \ref{tab:Results_data}). These luminosities $\langle v_{\textup{eje}}\rangle \sim 0.2$ are broadly consistent with those measured for kilonovae in general \cite{Ascenzi:2018mbh,Rastinejad:2021nev} and the $L_{\textup{knova}}\sim 10^{41}$erg s$^{-1}$ reported for kilonova associated with GRB 170817A \cite{Villar:2017wcc}, while the ejecta masses and $\tau_{\textup{peak}}$ are slightly larger than those inferred from observations. The temperature of $T_{\textup{peak}} \sim 1300$K corresponds to a peak wavelength of $\lambda_{\textup{peak}} = 2.22 \times 10^3 \textup{nm}(T_{\textup{peak}}/10^{3.11})$ with a range of $\lambda_{\textup{peak}} \sim 1907-2240$nm for the different cases. 

\subsection{Magnetic energy amplification}
\label{sec:ME_amp}

\begin{figure}[h]
    \centering
\includegraphics{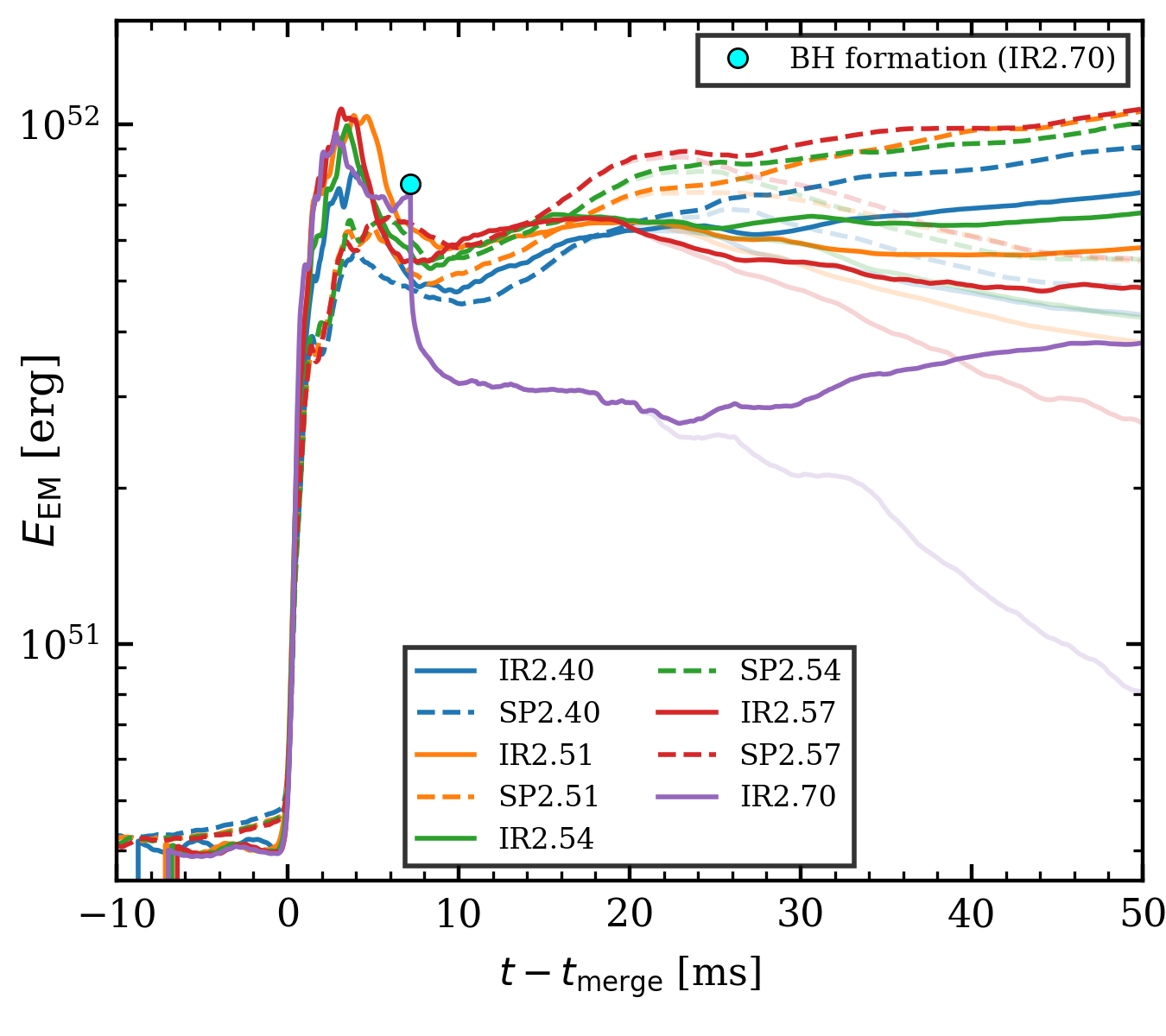}
    \caption{Total magnetic energy $E_{\textup{EM}}$ including both the energy contained in the simulation box, $E_{\textup{EM}}^{\rm in}$, plus the energy lost from the box via EM radiation. The $E_{\textup{EM}}^{\rm in}$ on its own is also shown using faded-out lines. The blue circle denotes the time of BH formation for the IR2.70 case, the only one which forms a BH by the time our simulations terminate.}
    \label{fig:EMF_energy}
\end{figure}

\begin{figure}
\begin{tabular}{cc}

\includegraphics{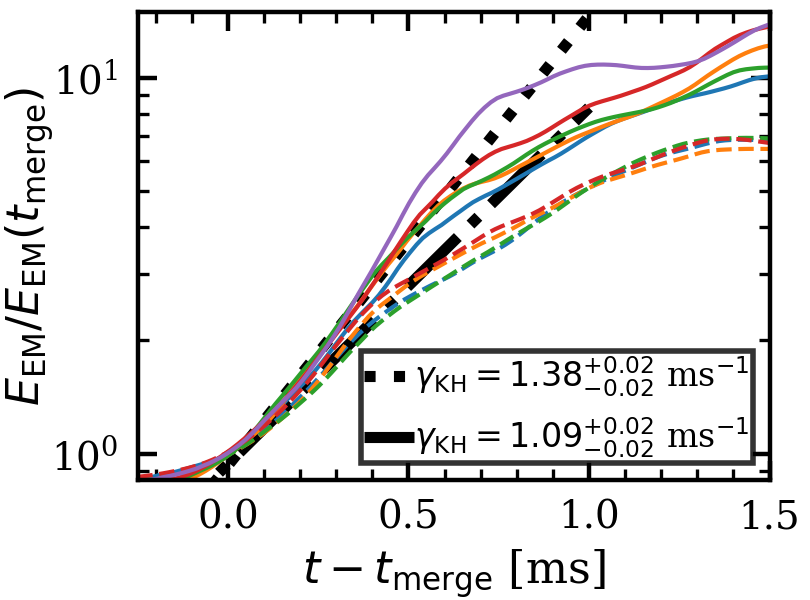} & \includegraphics{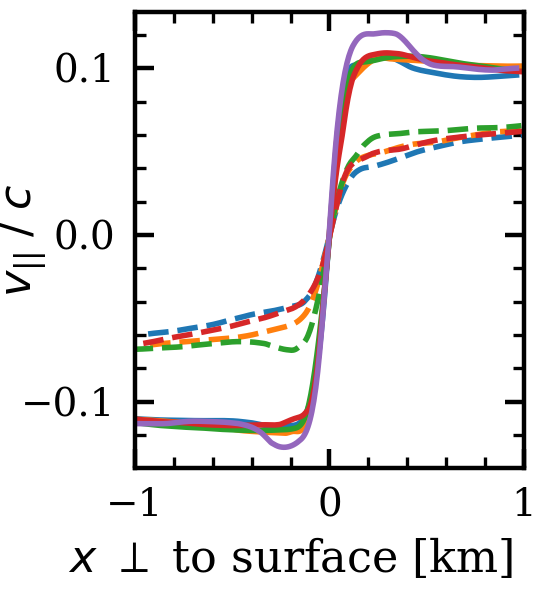} 
\end{tabular}
\caption{Left panel: the total magnetic energy in the first 1.5ms post-merger, normalised by the value at $t = t_{\textup{merge}}$, along with two exponential $\propto \exp(2\gamma_{\textup{KH}}t)$ trendlines fit to the irrotational and spinning cases, respectively. Right panel: the velocity parallel to the shear surface, $v_{\vert\vert}$, along a line perpendicular to the shear surface at $t \approx t_{\textup{merge}}$. Both plots use the same key for the different cases as in Fig. \ref{fig:EMF_energy}.}
\label{fig:KHI_amplification}
\end{figure}

The total magnetic energy, including the EM outflow from the simulation box, is shown in Fig. \ref{fig:EMF_energy}. Immediately after merger the magnetic energy increases by over an order of magnitude within a few ms, consistent with amplification via the KHI instability at the shear surface \cite{Kiuchi:2015sga} (with KH vortices visible). The relative amplification in the first 1.5ms is shown in Fig. \ref{fig:KHI_amplification} (left panel) with the black lines indicating fits to the initial linear regime for the irrotational (IR) and spinning (SP) cases respectively. The SP cases exhibit slower amplification ($\gamma_{\textup{KH}} \sim 1.09$) compared to the irrotational cases ($\gamma_{\textup{KH}} \sim 1.38$), which we can attribute to the reduced speed difference across the shear surface (shown in the right panel of Fig. \ref{fig:KHI_amplification}). Note however that the magnetic energy growth rate $2\gamma_{\textup{KH}} \sim 2\textup{ms}^{-1}$ is several orders of magnitude lower than the expected instability growth rate from linear perturbation theory of $\sigma_{\textup{KH}} \sim 10^2 \textup{ms}^{-1}$ from $\Delta v \sim 0.1c, d \sim 400\rm m$. After the KHI growth terminates at $\sim 2-6$ms, with a maximum magnetic field strength of $\sim 10^{17}$G, the magnetic energy falls. This decline (also seen in higher resolution simulations \cite{Palenzuela:2021gdo,Aguilera-Miret:2023qih}) may be due to the conversion of magnetic to kinetic energy in accelerating the ejecta or to the collapse of an unstable magnetic configuration. However, further study is needed to precisely determine the dynamics. 

The magnetic energy also declines sharply on BH formation for the IR2.70 case as the highly magnetized core of the star is swallowed by the BH, with a further slower decline up to $\sim 20$ms post-merger, as material continues to be accreted. Then all cases experience a growth in the total magnetic energy due to magnetic winding and the MRI, balanced by the loss of magnetic energy to the kinetic energy of the accelerated gas (particularly apparent for the IR2.57 case with the highest baryon pollution where the total magnetic energy, including the EM energy lost from the simulation box, continues to decline). We find the Shakura–Sunyaev parameter $\alpha_{\textup{SS}}$ in this regime is order $10^{-4}-10^{-3}$ in the core of the remnant stars, increasing to $\sim 10^{-2}$ in the outermost parts, although we note this parameter depends on the resolution \cite{Kiuchi:2017zzg}. The spinning cases also show a noticeably larger growth rate in this regime, potentially due to the larger initial total angular momentum. 

 \begin{figure*}[th]
    \subfloat{
        \centering
    \includegraphics[width=0.5\textwidth]{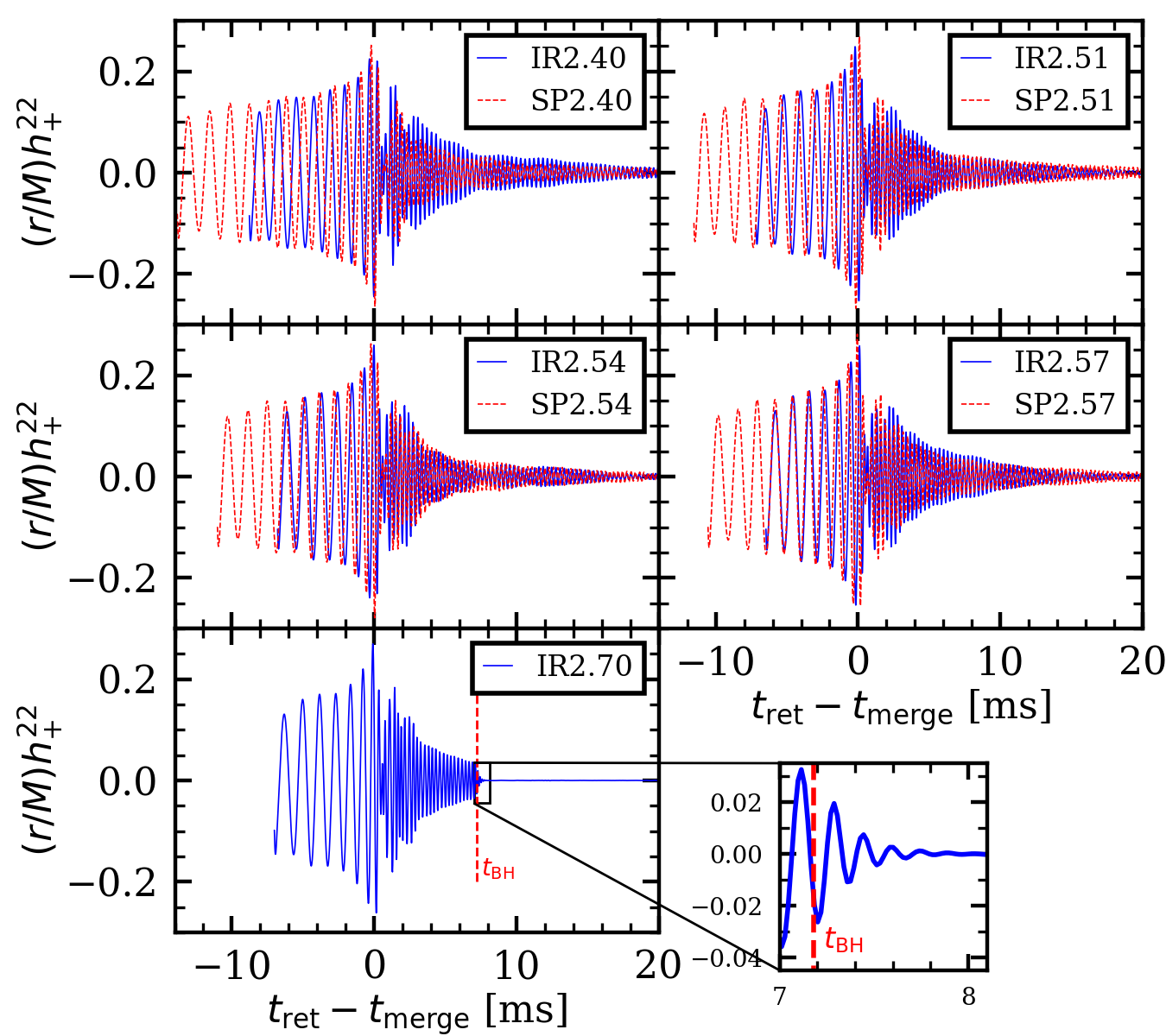}
    }
    \subfloat{
        \centering
    \includegraphics[width=0.50\textwidth]{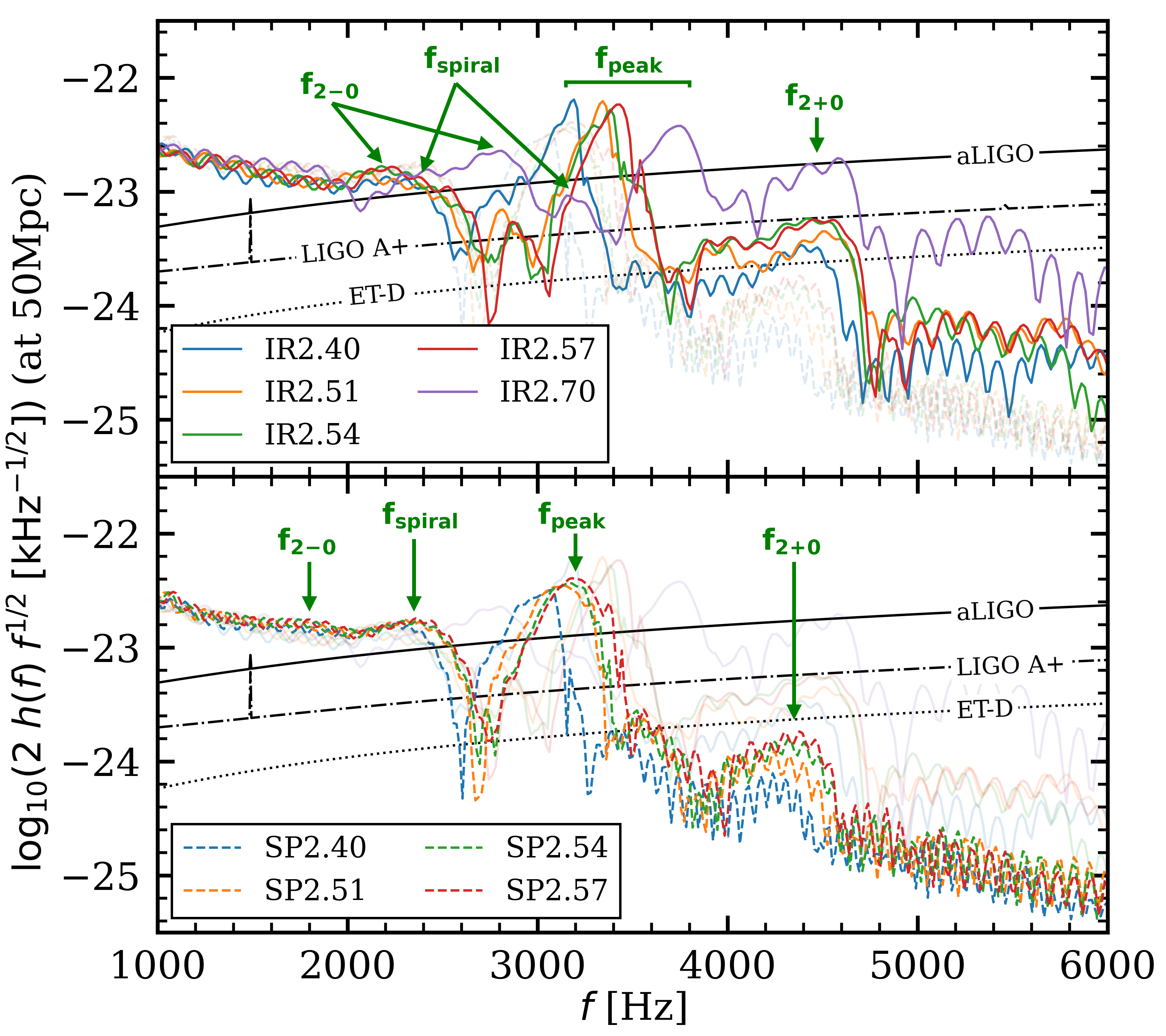}
    }
    \caption{Left: Gravitational wave strain $h^{22}_{+}$ (the 22 is the dominant mode) as a function of time measured from the peak of the GW signal at $t_{\textup{merge}}$, extracted at coordinate radius $r_{\textup{ext}} = 683\textup{km} \approx 180M$. The ringdown portion of the signal for the IR2.70 case that forms a BH is shown as a magnified insert. Right: The GW power spectrum of the dominant mode, for a source at distance 50Mpc, with the irrotational cases highlighted on the top right plot and the spinning cases highlighted on the bottom right plot. We also show the noise curves for aLIGO \cite{aLIGO:2020wna}, LIGO A+ \cite{LIGO:2018Aplus} and the Einstein Telescope in the D configuration (denoted ET-D) \cite{Hild:2010id} and the approximate locations of the major frequency peaks $f_{2\pm0},f_{\textup{spiral}},f_{\textup{peak}}$.}
    \label{fig:GWstrain}
\end{figure*}

\subsection{Gravitational waves}
\label{sec:GW}

The gravitational wave strain for the dominant $(2,2)$ mode and $h_{+}$ polarizations vs retarded time post-merger for the different cases is shown in the left panel of Fig. \ref{fig:GWstrain}. We see that prior to merger the waveforms of the spinning and irrotational cases match closely, apart from a slight dephasing and the longer inspiral for the spinning cases. After merger we see a higher frequency signal from the oscillations of the nonaxisymmetric remnants, which either slowly decay to zero as the stars settle down for the supramassive cases, or transition to the familiar BH ringdown signal (shown in an insert) for the IR2.70 hypermassive case which collapses to a BH. The amplitude of the post-merger signal is decreased for the spinning compared to the irrotational cases, with a difference of $\sim 50\%$ for the lowest mass $2.40 M_{\odot}$ cases vs $\sim 30\%$ for the $2.57 M_{\odot}$ cases. 

\begin{figure}[h]
    \centering
    \includegraphics{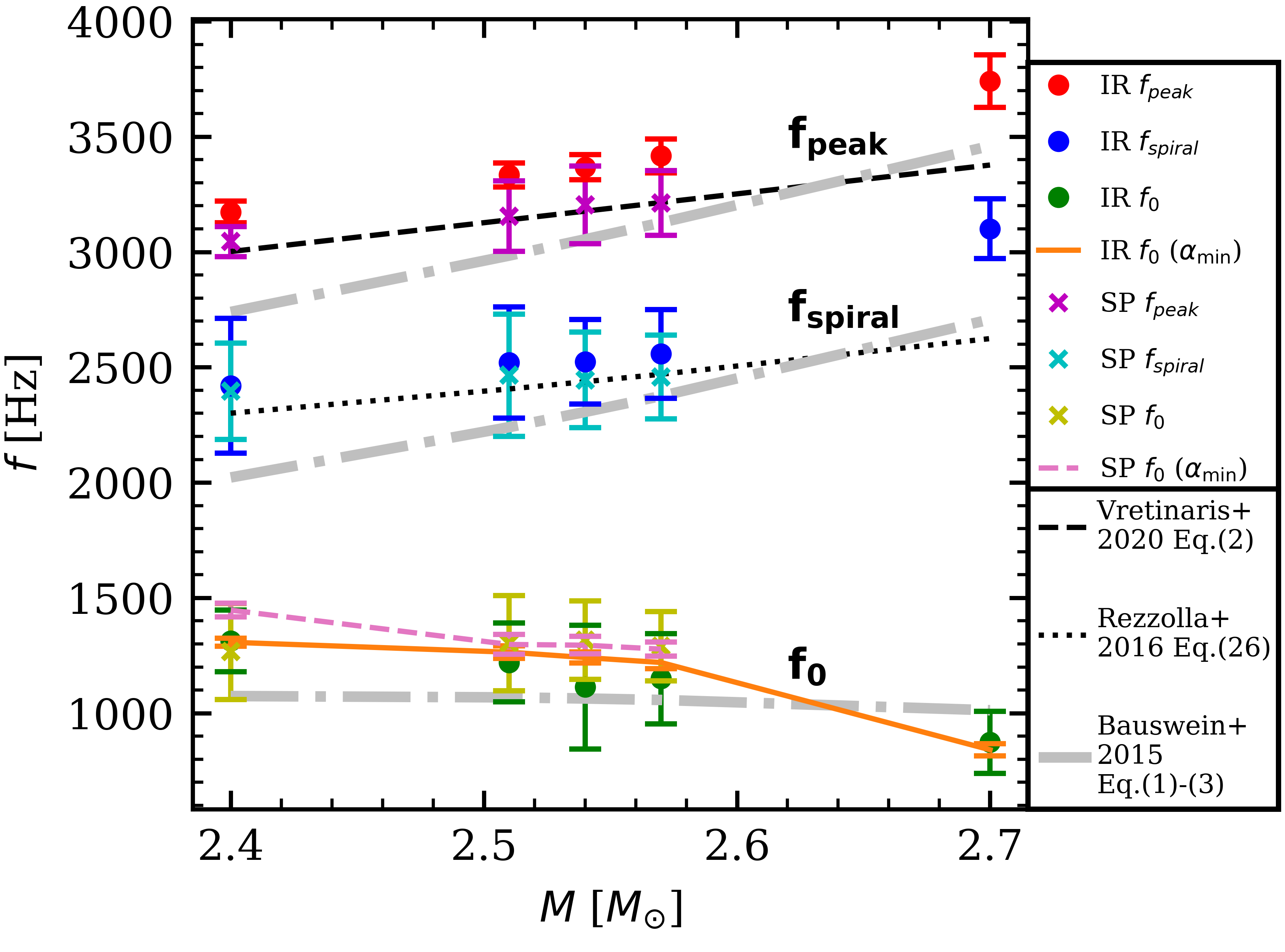}
    \caption{Characteristic frequencies of the post-merger GW signal for the different NS remnants vs initial total binary gravitational mass. Error bars denote uncertainties taken from the width of the fitted Gaussian peaks. The solid orange and dashed pink lines denote the $f_0$ quasi-radial frequency, as measured by fitting a Lorentzian peak to the Fourier spectrum of the oscillations of the minimum lapse $\alpha_{\textup{min}}$, for the irrotational cases and spinning cases respectively. Also plotted are empirical quasi-universal relations given in several previous GRHD (general relativistic hydrodynamics) numerical studies without magnetic fields.}
    \label{fig:GW_f_peaks}
\end{figure}

The power spectrum of the post-merger signal is shown in the right panel of Fig. \ref{fig:GWstrain}. As in our previous studies \cite{Ruiz:2021qmm} and other numerical works (see for example \cite{Stergioulas:2011gd,Takami:2014tva,Bauswein:2015yca,Rezzolla:2016nxn,Vretinaris:2019spn,Raithel:2022orm}) we can identify four characteristic peaks in the post-merger power spectrum, denoted (following \cite{Stergioulas:2011gd,Bauswein:2015yca}) $f_{\textup{peak}},f_{2\pm 0},f_{\textup{spiral}}$. The $f_{\textup{peak}} = \Omega/\pi$ is the largest amplitude frequency mode, corresponding to the quadrupolar $l=m=2$ fundamental fluid mode in the remnant. The $f_{\textup{spiral}} = \Omega_{\textup{spiral}}/\pi$ peak corresponds to the rotation of the two bulges of matter that form at the surface of the remnant, making a rotating two-armed spiral pattern \cite{Bauswein:2015yca} with angular frequency $\Omega_{\textup{spiral}}$. This rotates more slowly than the inner remnant core, with $\Omega_{\textup{spiral}} < \Omega$, and therefore $f_{\textup{spiral}} < f_{\textup{peak}}$. Finally, the $f_{2\pm 0}$ peaks result from nonlinear interaction between the $l=m=2$ quadrupolar mode and the $m=0$ fundamental quasi-radial oscillation mode with frequency $f_0$, with $f_{2\pm 0} \approx f_{\textup{peak}} \pm f_0$. 

To measure the characteristic frequencies we fit Gaussian peaks to the spectrum of the signal from $t=t_{\textup{start}} = t_{\textup{merge}}-0.25(t_{\textup{end}}-t_{\textup{merge}})\delta/(2-\delta)$ to $t=t_{\textup{end}}$, the end of the simulation, where $\delta=0.05$ is the parameter of the Tukey window function used. The results for $f_{\textup{peak}},f_{\textup{spiral}}$ and $f_{0} = (f_{2+0}-f_{2-0})/2$ are shown in Fig. \ref{fig:GW_f_peaks}, along with lines denoting the empirical relations derived from numerical GRHD (GR-hydrodynamics without magnetic fields) simulations: specifically Vretinaris et al. (2020) \cite{Vretinaris:2019spn} Eq. (2), Rezzolla \& Takami (2016) \cite{Rezzolla:2016nxn} Eq. (26), and Bauswein \& Stergioulas (2015) \cite{Bauswein:2015yca} Eqs. (1)-(3). For the spinning (SP) cases we see good agreement between $f_{\textup{peak}}$ and the Vretinaris+ relation and $f_{\textup{spiral}}$ and Rezzolla+ relation. The irrotational (IR) $f_{\textup{peak}}$ values are some $\sim 200$Hz higher in frequency than those predicted by the empirical GRHD relations, with the frequency for the hypermassive IR2.70 $\gtrsim 300$Hz higher than the empirical relation predictions. This is consistent with the results reported in \cite{Ruiz:2021qmm} which showed that the presence of magnetic fields can alter the characteristic frequencies, shifting $f_{\textup{peak}}$ in particular to higher frequencies. Finally, we plot $f_0$ as directly measured from the dominant oscillation frequency of the minimum lapse $\alpha_{\textup{min}}$, which serves as a good proxy for oscillations at the center of the remnant \cite{Rezzolla:2016nxn,Bauswein:2015yca} and find good agreement with the values inferred from the GW spectrum. 

\section{Conclusions}
\label{sec:Conclusions}

GW events with electromagnetic counterparts, like GW170817 and GRB 170817A, are invaluable for addressing several open questions in fundamental physics. They can provide independent estimates of the expansion rate of the universe \cite{LIGOScientific:2017adf}, put constraints on the maximum mass, tidal deformability, and mass-radius relation of neutron stars and thus inform our understanding of their nuclear equation of state \cite{LIGOScientific:2017vwq,Margalit:2017dij,Ruiz:2017due}, and have provided unambiguous evidence that compact binary mergers containing NSs can be progenitors of the central engines that power sGRBs and associated kilonovae \cite{LIGOScientific:2017adf,Savchenko:2017ffs}. To make the most of the scientific potential of future multimessenger observations theoretical modelling is crucial. Here we sought to examine one of the key open questions: what are the central engines for sGRBs and the ultrarelativistic jets thought to produce the $\gamma-$ray emission? In previous works we conducted some of the first self-consistent GRMHD simulations demonstrating the formation of an incipient jet from BHNS mergers \cite{Paschalidis:2014qra,Ruiz:2018wah} and from NSNS mergers which result in a metastable HMNS which undergoes delayed collapse to a BH \cite{Ruiz:2016rai,Ruiz:2017inq,Ruiz:2019ezy,Ruiz:2020via,Ruiz:2021qmm}. 

In this study we extended our work by performing a set of self-consistent GRMHD simulations of NSNS mergers that result in long-lived SMNS remnants, along with a benchmark merger that produces a transient HMNS which collapses to a BH. For the benchmark HMNS case we observe the formation of a low density evacuated funnel (half-opening angle $\sim 30^{\circ}$) above the BH poles following BH formation which is magnetically dominated ($b^2/(2\rho_0) \gg 1$), collimated helical magnetic field lines extending from the poles, and mildly relativistic ($\Gamma \gtrsim 2.0$) outflow, with a EM luminosity ($\sim 10^{52}\textup{erg}\;\textup{s}^{-1}$ by the end of the simulation) consistent with the Blandford-Znajek (BZ) mechanism \cite{Thorne:1986} and an accretion efficiency consistent with a Novikov-Thorne accretion disk \cite{Shapiro:2004ud}. In agreement with our previous studies we identify the outflow as meeting the criteria for an incipient jet. 

For the SMNS cases we also observe the formation of a partially evacuated funnel (half-opening angle $\sim 25^{\circ}$) above the poles of the NS remnants. We observe a helical magnetic field within this funnel region that is actually stronger than in the HMNS/BH case, a mildly relativistic outflow, and find that the SMNS cases produce a high fluid and EM luminosity ($\sim 10^{53}\textup{erg}\;\textup{s}^{-1}$) for a longer duration than the HMNS case. From the ejecta mass ($M_{\textup{esc}} \gtrsim 10^{-2} M_{\odot}$) and average ejecta velocity ($\langle v_{\textup{eje}} \rangle \sim 0.2c$) we estimate the bolometric luminosity for the associated kilonova as $L_{\textup{knova}} \sim 10^{41} \textup{erg}\;\textup{s}^{-1}$, broadly consistent with observations \cite{Ascenzi:2018mbh,Rastinejad:2021nev,Villar:2017wcc}. The EM luminosity may be consistent with the higher end of observed $\gamma-$ray luminosities from sGRBs \cite{Li:2016pes,Beniamini:2020adb}, or in excess, depending on uncertainties about the $\gamma-$ray prompt emission mechanism and the efficiency of generating $\gamma-$ray photons. 

The gas density in the funnel is also significantly larger compared to the BH case, and this baryon pollution means the outflow is less magnetically dominated (i.e. $b^2/(2\rho_0)$ is smaller) limiting the magnetic energy per unit mass available for accelerating the gas to ultrarelativistic speeds. 

In addition, we note that for outflows from magnetized NS remnants we do not have the same kind of evidence for efficient acceleration and ultrarelativistic jet formation that we have for BZ-powered jets from BHs with magnetized accretion disks. Our results suggest one cannot simply assume that magnetic energy at the base of the outflow is efficiently converted to kinetic energy asymptotically, as for idealized semi-analytic models of ideal MHD jets, and that in real dynamical system with a large amount of baryon loading significant amounts of energy may be lost via mixing between the low density funnel and the higher density debris torus (which may also reduce the $\gamma-$ray conversion efficiency). However, we emphasise that further simulations on larger spatial scales, akin to e.g. \cite{Gottlieb:2022sis}, are needed to fully resolve this question. Therefore while we note that we observe \textit{jet-like structures} in the outflow from the SMNS cases which meet our essential criteria for an \textit{incipient} jet, we cannot conclude they will actually produce a true ultrarelativistic jet as required for the observed $\gamma-$ray emission or correspond to the central engines for sGRBs. 

We also investigated the effect of the initial mass and spin of the NSs. The mass of the SMNS remnant has only a limited effect on the post-merger outflow. The higher the mass of the remnant the more compact the bound torus of debris, and the highest mass irrotational SMNS case showed noticeably more baryon pollution and a lower luminosity than the other cases. However, there was little difference between the baryon pollution of luminosity between highest and lowest mass SMNS cases with initial NS spin, or between the other irrotational cases. There are more consistent differences between the spinning and irrotational cases. The growth rate of the magnetic field due the KHI instability in the first few ms post-merger is lower for the spinning cases than the irrotational due to the lower velocity difference across the initial shear surface, while at later times the growth rate due to the MRI instability and magnetic winding is greater for the spinning cases due to the larger total initial angular momentum. The EM luminosity and ejecta mass is also larger compared to the irrotational cases of the same mass. Finally, we note see that the high frequency post-merger component of the GW signal is smaller for the spinning cases, although for a merger at a typical distance of 50Mpc this difference may be undetectable with current and future GW observatories. 

Our simulations do not include neutrino radiation transport,
although our preliminary study \cite{Sun:2022vri} suggests that our main results will still hold even when neutrinos
are present. Further studies are needed that will extend the current investigation and provide a definite answer.

Movies and additional 3D visualizations highlighting
our simulations can be found at \cite{website}.

\section*{Acknowledgements}
We thank members of our Illinois Relativity Undergraduate Research Team (Nawaf Aldrees, Jonah  Doppelt, Rohan Narasimhan, Yinuan Liang, Eric Yu) for assistance with some of the 3D visualizations. This work was supported in part by National Science Foundation (NSF) Grants No. PHY-2308242, No. OAC-2310548 and No. PHY-2006066 to the University of Illinois at Urbana-Champaign. 
M.R. acknowledges support by the Generalitat Valenciana Grant CIDEGENT/2021/046 and by the Spanish Agencia Estatal de Investigaci\'on (Grant PID2021-125485NB-C21). 
A.T. acknowledges support from the National Center for Supercomputing Applications (NCSA) at the University of Illinois at Urbana-Champaign through the NCSA Fellows program. 
This work used Stampede2 at TACC and Anvil at Purdue University through allocation MCA99S008,
from the Advanced Cyberinfrastructure Coordination Ecosystem: Services \& Support (ACCESS) program, which is supported by National Science Foundation grants \#2138259, \#2138286, \#2138307, \#2137603, and \#2138296.
This research also used Frontera at TACC through allocation AST20025. Frontera is made possible by NSF award OAC-1818253. 

The authors thankfully acknowledge the computer resources at MareNostrum and the technical support provided by the Barcelona Supercomputing Center (AECT-2023-1-0006).
 
\bibliographystyle{apsrev4-1}
\bibliography{apssamp}
\end{document}